\documentclass[12pt]{iopart_mod}

\usepackage{amsmath,amssymb,amsfonts}
\usepackage{mathrsfs,cite,yfonts}
\usepackage{picture,epic,graphicx,color,slashed} 
\newcommand{\beqs}{\begin{equation*}}
\newcommand{\beq}{\begin{equation}}

\newcommand{\eeqs}{\end{equation*}}
\newcommand{\eeq}{\end{equation}}

\newcommand{\beqas}{\begin{eqnarray*}}
\newcommand{\beqa}{\begin{eqnarray}}

\newcommand{\eeqas}{\end{eqnarray*}}
\newcommand{\eeqa}{\end{eqnarray}}

\newcommand{\eq}[2]{\begin{equation} #1 \label{#2} \end{equation}}

\newcommand{\eps}{\varepsilon}
\newcommand{\al}{\alpha}
\newcommand{\be}{\beta}
\newcommand{\ga}{\gamma}
\newcommand{\de}{\delta}

\newcommand{\la}{\lambda}
\newcommand{\si}{\sigma}

\newcommand{\Ga}{\Gamma}

\newcommand{\blist}{\begin{itemize}}
\newcommand{\elist}{\end{itemize}}

\DeclareMathOperator{\extdm}{d}
\newcommand{\extd}{\extdm \!}

\providecommand{\href}[2]{#2}

\begin{document}

\title{Holographic applications of logarithmic conformal field theories}
\author{D. Grumiller, W. Riedler, J. Rosseel}

\address{Institute for Theoretical Physics,
           Vienna University of Technology,\\
           Wiedner Hauptstrasse 8-10/136,
           Vienna, A-1040, Austria 
} 

\author{T. Zojer}

\address{Centre for Theoretical Physics, University of Groningen,\\
           Nijenborgh 4, 9747 AG Groningen, The Netherlands
}

\eads{\mailto{grumil@hep.itp.tuwien.ac.at}, \mailto{riedler@hep.itp.tuwien.ac.at}, \mailto{rosseelj@hep.itp.tuwien.ac.at}, \mailto{t.zojer@rug.nl}}


\begin{abstract}
We review the relations between Jordan cells in various branches of physics, ranging from quantum mechanics to massive gravity theories. Our main focus is on holographic correspondences between critically tuned gravity theories in Anti-de~Sitter space and logarithmic conformal field theories in various dimensions. We summarize the developments in the past five years, include some novel generalizations and provide an outlook on possible future developments.
\end{abstract}



\section{Introduction}\label{se:1}

The Anti-de~Sitter/Conformal Field Theory (AdS/CFT) correspondence \cite{Maldacena:1997re} provides a realization of the holographic principle \cite{'tHooft:1993gx,Susskind:1995vu} within string theory. Originally, the holographic principle was motivated by the desire to have a unitary quantum theory of gravity. However, the way the AdS/CFT correspondence works makes it plausible that holography may also apply if neither the gravity theory nor the field theory is unitary. 

Non-unitarity is often automatically associated with some sickness of the theory. Frequently this is the correct association. However, there are many systems where non-unitarity is not a bug, but a feature. This includes, for instance, open quantum systems and systems with (quenched) disorder. The kind of non-unitarity we are interested in arises in a specific way through the emergence of Jordan cells in the Hamiltonian. Such Jordan cells have been observed experimentally through Berry phases in non-hermitian quantum mechanics and are the defining property of logarithmic CFTs (log CFTs), see \cite{Gurarie:1993xq} and the review articles \cite{Flohr:2001zs,Gaberdiel:2001tr} (additionally, see the other contributions in this Special Issue and references therein, like \cite{Cardy:2013rqg}). About five years ago it was realized that such Jordan cells arise on the gravity side \cite{Grumiller:2008qz} for a certain critical tuning of the coupling constants \cite{Li:2008dq} in topologically massive gravity (TMG) \cite{Deser:1982vy,Deser:1982wh,Deser:1982a}. 

The discovery of Jordan cells on the gravity side and the ensuing holographic correspondence to log CFTs has led to many interesting developments. The main purpose of this review article is to provide an introduction to the emergence of Jordan cells and an overview of the evolution of the AdS/log CFT correspondence in the past five years.
This paper is organized as follows.
\begin{itemize}
 \item In the remainder of this section we introduce Jordan cells and their importance in quantum mechanics in section \ref{se:1.1}. Jordan cells are the defining property of log CFTs, whose basic features we summarize in section \ref{se:1.2}. Section \ref{se:1.3} provides a crash course on the AdS$_3$/CFT$_2$ correspondence, with no strings or logs attached.
 \item In section \ref{se:2} we review the AdS$_3$/log CFT$_2$ correspondence. We start with critical TMG in section \ref{se:2.1} and collect the checks of this correspondence in section \ref{se:2.2}. Section \ref{se:2.3} provides results on fairly generic critical massive gravity theories.
 \item In section \ref{se:3} we consider various generalizations and include some novel material. We start with a supersymmetric version of the AdS$_3$/log CFT$_2$ correspondence in section \ref{se:3.0}. Then we move to higher dimensions and summarize the status of the AdS$_{d+1}$/log CFT$_d$ correspondence in section \ref{se:3.1}. In section \ref{se:3.2} we outline possible generalizations to other spins (including higher spin gravity) and backgrounds that differ from AdS. In section \ref{se:3.3} we study \.In\"on\"u--Wigner contractions that lead to log Galilean Conformal Algebras (log GCAs).
 \item In section \ref{se:4} we collect some loose ends. Remarks on condensed matter applications are provided in section \ref{se:4.2}.  Missing checks are listed in section \ref{se:4.1}. Finally, section \ref{se:4.3} speculates on some open issues.
\end{itemize}

\subsection{Jordan cells in quantum mechanics}\label{se:1.1}

Algebra is ubiquitous in Mathematics and Physics. A relevant aspect of linear algebra is matrices and their eigenvalues. If for all eigenvalues the geometric multiplicities equal the algebraic multiplicities then the matrix is diagonalizable and the number of eigenvectors equals to the rank of the matrix. For ``critical tunings'' of the matrix entries it may happen that some geometric multiplicities are smaller than the algebraic ones. Then the matrix is not diagonalizable but instead has at least one non-trivial Jordan block. In that case the number of eigenvectors is smaller than the rank of the matrix and generalized eigenvectors emerge. We shall refer to non-trivial Jordan blocks as ``Jordan cells''.

Before explaining some physics implications of critical tunings that lead to Jordan cells let us present a simple example of the situation described in the previous paragraph. Take the $2\times 2$ matrix $H^{\rm crit}$,
\eq{
H^{\rm crit}=\left(\begin{array}{cc}
       \lambda & 1 \\ 0 & \lambda
      \end{array}\right)\,,
}{eq:lcft1}
with eigenvalue $\lambda$, whose algebraic multiplicity is 2, while its geometric multiplicity is 1. The matrix $H^{\rm crit}$ is not diagonalizable; in fact, $H^{\rm crit}$ is already presented in Jordan normal form and has a Jordan cell of rank 2. (We call it ``rank 2'' even for $\la=0$, where the matrix rank is reduced by one.) The eigenvector $\psi$ (up to normalization) is given by
\eq{
\psi = \left(\begin{array}{c}
                                        1 \\ 0
                                       \end{array}\right)
\qquad (H^{\rm crit}-\lambda)\psi = 0 \,.
}{eq:lcft2}
The generalized eigenvector $\psi^{\rm log}$ (again up to normalization) is given by
\eq{
\psi^{\rm log} = \left(\begin{array}{c}
                                        0 \\ -1
                                       \end{array}\right)
\qquad (H^{\rm crit}-\lambda)^2\psi^{\rm log} = 0 \qquad (H^{\rm crit}-\lambda)\psi^{\rm log} = \psi\,.
}{eq:lcft3}
The superscript ``log'' is chosen for later convenience; at the moment it is just a label to discriminate the generalized eigenvector $\psi^{\rm log}$ from the eigenvector $\psi$. The generalized eigenvector $\psi^{\rm log}$ is not defined uniquely; in fact, the linear combination $\tilde\psi^{\rm log}=\psi^{\rm log}+\gamma\psi$ is again a generalized eigenvector for any complex $\gamma$. 
It is possible to define a useful inner product with the matrix
\eq{
\eta=\begin{pmatrix} 0 & -1 \\ -1 & 1 \end{pmatrix}\,.
}{eq:lcft12}
We shall see later how this matrix emerges naturally.
The eigenvector $\psi$ is null with respect to the inner product $\langle\psi\psi\rangle=\psi^T \eta \psi$.
\eq{
\langle\psi\psi\rangle=0
}{eq:lcft5}
The inner product between eigenvector and generalized eigenvector is non-zero and independent from $\gamma$.
\eq{
\langle \psi^{\rm log} \, \psi \rangle = \langle \tilde \psi^{\rm log} \, \psi \rangle = 1
}{eq:lcft6}
(The inner product between the generalized eigenvector with itself does depend on $\gamma$.)

It is straightforward to generalize the discussion above to higher rank Jordan cells. For a rank-$n$ Jordan cell,
\eq{
J^{(n)} = \left(\begin{array}{ccccc}
                      \la & 1 & 0 & \cdots & 0 \\
                      0   & \la & 1 & \ddots & 0 \\
                      \vdots & \ddots & \ddots & \ddots & \vdots \\
                      0 & \cdots & 0 & \la & 1 \\
                      0 & \cdots & 0 & 0 & \la
                     \end{array}\right)
}{eq:lcft7}
there is a single eigenvector $\psi=(1, \,0, \,\dots, \,0)^T$ and $n-1$ generalized eigenvectors,
\eq{
\big(J^{(n)}-\la\big)^m\psi^{\textrm{log}(m-1)} = 0 \qquad \big(J^{(n)}-\la\big)^{m-1}\psi^{\textrm{log}(m-1)} = \psi^{\textrm{log}(m-2)}\qquad m=2, \dots, n
}{eq:lcft8}
where $\psi^{\textrm{log}(0)}=\psi$. The generalized eigenvectors $\psi^{\textrm{log}(m-1)}$ are defined only up to arbitrary additions of $\psi^{\textrm{log}(k-1)}$ with $k<m$.

We consider now the probably simplest physics application of Jordan cells in the context of non-hermitian quantum mechanics (see e.g.~\cite{Moiseyev:2011}, in particular chapter 9). Suppose we have the complex $2\times 2$ Hamiltonian
\eq{
H=\left(\begin{array}{cc}
       \lambda+1 & \mu \\ \mu & \lambda-1
      \end{array}\right)\,,
}{eq:lcft4}
with some real parameter $\lambda$ and a complex parameter $\mu$. We assume here that these parameters can be adjusted arbitrarily, e.g.~by setting up an experiment appropriately or by tuning some effective coupling constant suitably. The eigenvalues of $H$ are given by $E_{\pm} = \la\pm\sqrt{1+\mu^2}$. For generic values of the parameter $\mu$ the eigenvalues differ from each other and thus the Hamiltonian must be diagonalizable. The only exception arises for the critical tunings $\mu^{\rm crit}=\pm i$, where both eigenvalues coalesce, $E_+=E_-=\lambda$.\footnote{The coalescence of the energy eigenvalues is not at odds with the non-crossing rule by Neumann--Wigner \cite{Neumann:1929} and Hund \cite{Hund:1927}, since hermiticity is assumed in the derivation of this rule, whereas the Hamiltonian \eqref{eq:lcft4} necessarily becomes non-hermitian for $\mu^2=-1$.} Taking with no loss of generality the solution $\mu^{\rm crit}=-i$ and making a similarity transformation $S^{-1}HS$ to bring the Hamiltonian into Jordan normal form then leads to the matrix $H^{\rm crit}$ in \eqref{eq:lcft1}. The transformation matrix reads
\eq{
S=\left(\begin{array}{cc}
   i & 0 \\ 1 & -1
  \end{array}\right)\,.
}{eq:lcft11}
We can thus take all the results from the previous two paragraphs and translate them into the language of non-hermitian quantum mechanics: the critical tuning leading to a Jordan cell \eqref{eq:lcft1} is called ``exceptional point'' or ``branch point'' (see e.g.~\cite{Klaiman:2008zz}); the fact that the eigenvector \eqref{eq:lcft2} becomes null \eqref{eq:lcft5} is called ``self-orthogonality''; the existence of a generalized eigenvector \eqref{eq:lcft3} is called ``incompleteness of the spectrum''.

For the phenomenological importance of Jordan cells in non-hermitian quantum mechanics we refer to the textbook by Moiseyev \cite{Moiseyev:2011} and references therein. We just mention one relevant aspect here. Namely, the fact that Jordan cells require an infinite fine-tuning of the parameters implies that in experiments or numerical simulations one never sits exactly at the critical/exceptional point, but only in some vicinity thereof. Therefore, one has to find observables that are accessible also slightly away from criticality. An example of such observables are Berry phases, see \cite{Moiseyev:2011} and references therein.

It is often useful to consider criticality as a limit of near-critical configurations. We explain now how this works. For $\mu^2\neq -1$ the two eigenvectors of $H$ in \eqref{eq:lcft4} are given by
\eq{
\psi_1 = \left(\begin{array}{c} 
                \frac{1 - \sqrt{1 + \mu^2}}{\mu} \\ 1
               \end{array}\right)
\qquad \psi_2 = \left(\begin{array}{c} 
                \frac{1 + \sqrt{1 + \mu^2}}{\mu} \\ 1
               \end{array}\right)\,.
}{eq:lcft9}
In the near critical limit $\mu\approx-i$ it is useful to define the small parameter $\eps:=\sqrt{1+\mu^2}$ such that $\lim_{\eps\to 0}\mu=\mu^{\rm crit}$. In that limit both eigenvectors approach the critical eigenvector $\psi=\lim_{\eps\to 0}[\psi_1(\eps)+\psi_2(\eps)]/2=(i,\, 1)^T$. [Note that this is precisely the transformation matrix $S$ from \eqref{eq:lcft11} acting on the eigenvector \eqref{eq:lcft2}.] However, there is a linearly independent combination of $\psi_{1,2}$ that survives the limit and differs from $\psi$. It is given by
\eq{
\psi^{\rm log} = \lim_{\eps\to 0} \frac{\psi_1(\eps) - \psi}{\eps} = \frac{\extd}{\extd\eps} \psi_1(\eps) \big|_{\eps=0} \,.
}{eq:lcft10}
The quantity $\psi^{\rm log}=(-i, \,0)^T$ is the generalized eigenvector. The inner products are given by $\langle \psi \psi\rangle=\psi^T \psi=0$ and $\langle\psi^{\rm log}\psi\rangle=1$. Consistently, we obtain $\eta=S^T S$, with $\eta$ from \eqref{eq:lcft12}, which explains why we made that choice.

Let us summarize key aspects of Jordan cells in quantum mechanics. 1.~Critical tuning of some parameters is required. 2.~Eigenvalues coalesce at the critical point. 3.~At least one generalized eigenvector $\psi^{\rm log}$ emerges. 4.~The eigenvector $\psi$ becomes null with respect to a specific inner product, but it has non-vanishing inner product with the generalized eigenvector $\psi^{\rm log}$. 5.~The generalized eigenvector $\psi^{\rm log}$ can be constructed through a limit of a suitable linear combination of eigenvectors away from criticality. These properties generalize to $1+1$ dimensional CFTs discussed in the next subsection.

\subsection{Log CFT basics}\label{se:1.2}

The prototypical example for a log CFT is the $c=-2$ triplet theory \cite{Gurarie:1993xq,Flohr:1995ea,Kausch:1995py} (see also \cite{Gaberdiel:1996np,Rohsiepe:1996qj,Flohr:2001zs,Gaberdiel:2001tr,Fjelstad:2002ei,Pearce:2006sz,Read:2007qq}).
In this review we are interested in log CFTs that arise as holographic duals of some gravity theories.
No gravity dual for the $c=-2$ model exists so far.
Therefore, we do not follow the historic route, but instead start with the kind of log CFTs that are useful for our purposes. 

Following the discussion in the previous subsection we consider a 1-parameter family of $1+1$-dimensional CFTs such that there is a critical tuning of this parameter where a Jordan cell emerges. We keep the construction as simple as possible; analogous constructions appeared e.g.~in \cite{Cardy:1999zp,Gurarie:1999yx,Cardy:2001,Kogan:2001ku,Rasmussen:2004gx,Rasmussen:2004na,Skenderis:2009nt,Grumiller:2010rm,Grumiller:2010tj,Cardy:2013rqg}.

As minimal amount of data we need the values of the central charges $c_L$, $c_R$. The minimal operator content consists of the (anti-)holomorphic flux components of the stress-energy tensor, $T_{zz}(z)=\psi_L(z)$, $\bar T_{\bar z\bar z}=\psi_R(\bar z)$ with conformal weights $(h_L, \,\bar h_L)=(2, \,0)$ and $(h_R, \,\bar h_R)=(0, \,2)$, respectively.\footnote{%
We refer to them also as left-(right-)moving flux components of the stress-energy tensor.
} 
This is not quite enough to obtain a Jordan cell in a CFT. Therefore, suppose there is some additional operator $\psi_\eps (z,\,\bar z)$ with conformal weights $(h_\eps, \,\bar h_\eps)=(2+\eps, \,\eps)$, where $\eps$ is some real parameter that can be varied. Suppose further that $c_L$ is a linear function of $\eps$, $c_L=-b_L \,\eps$, with some real constant $b_L\neq 0$. Then we have all the ingredients for the emergence of a Jordan cell.

Observe that for $\eps\to 0$ the operator $\psi_\eps$ degenerates with the operator $\psi_L$. This is very much like the coalescence of two eigenvectors into a single one. The limiting construction \eqref{eq:lcft10} suggests to define an operator
\eq{
\psi^{\rm log}(z, \,\bar z)= \lim_{\eps\to 0} \frac{\psi_\eps(z, \,\bar z) - \psi_L(z)}{\eps} = \frac{\extd}{\extd\eps}\psi_\eps(z, \,\bar z)\big|_{\eps=0}\,.
}{eq:lcft13}
The correlators with the operator $\psi^{\rm log}$ can then be calculated as a limit, using only known correlators. Let us focus here on 2-point correlators. For finite $\eps$ the non-vanishing ones are given by 
\begin{align}
 \langle\psi^L(z)\psi^L(0)\rangle &= \frac{c_L}{2z^4} \\
 \langle\psi^R(\bar z)\psi^R(0)\rangle &= \frac{c_R}{2\bar z^4} \\
 \langle\psi_\eps(z, \bar z)\psi_\eps(0, 0)\rangle &= \frac{c_\eps}{2z^{4+2\eps}\bar z^{2\eps}} \,.
\end{align}
Apart from the left and right central charges $c_{L,R}$ the 2-point correlators involve the normalization constant $c_\eps = - c_L +{\cal O}(\eps^2)$.

Correlators of the operators $\psi^{\rm log}$ and $\psi^L$ are then determined as simple limits $\eps\to 0$, exploiting $\langle\psi_\eps\psi^L\rangle=0$ for any finite $\eps$.
\begin{align}
  \langle\psi^L(z)\psi^L(0)\rangle &= 0 \label{eq:lcftc1} \\
 \langle\psi^{\rm log}(z, \bar z)\psi^L(0)\rangle &= -\lim_{\eps\to 0} \frac{\langle\psi^L(z)\psi^L(0)\rangle}{\eps} = \frac{b_L}{2z^4} \label{eq:lcftc2} \\
 \langle\psi^{\rm log}(z, \bar z)\psi^{\rm log}(0, 0)\rangle &= \lim_{\eps\to 0} \frac{\langle\psi_\eps(z, \bar z)\psi_\eps(0, 0)\rangle + \langle\psi^L(z)\psi^L(0)\rangle}{\eps^2} = -\frac{b_L\ln{(m_L^2|z|^2)}}{z^4} \label{eq:lcftc3}
\end{align}
The first correlator is reminiscent of the ``self-orthogonality'' in non-hermitian quantum mechanics \eqref{eq:lcft5}. Similarly, the second correlator resembles the inner product \eqref{eq:lcft6} and involves the constant $b_L$, which is sometimes called ``new anomaly''.\footnote{%
Since the overall factor in $\psi^L$ is fixed at finite $\eps$ the new anomaly $b_L$ is well-defined and independent from the normalization of $\psi^{\textrm{log}}$ once a choice has been made about the factors on the right-hand sides of the correlators \eqref{eq:lcftc2} and \eqref{eq:lcftc3}.
Note that a rescaling of $\psi^{\textrm{log}}$ would change these factors.
}
The third correlator is the most noteworthy one and the namesake of this kind of CFT. 
Its right hand side is proportional to the new anomaly and involves a logarithm of $|z|^2$.
This explains why we labeled this operator with the superscript ``log''.\footnote{
In this review we take an algebraic perspective, which is why we have started with Jordan cells in section  \ref{se:1.1}.
Another way to describe the essence of log CFTs is in terms of linear second order differential equations at a critical point, meaning that the two roots of the indicial equation coincide or differ by an integer and a logarithmic solution appears.
This was essentially the way log CFTs were introduced by Gurarie \cite{Gurarie:1993xq}.
} 
The quantity $m_L$ is a spurious mass parameter that depends on the numerical value of the ${\cal O}(\eps^2)$ term in $c_\eps$. It can be changed to any non-zero value by the redefinition $\tilde\psi^{\rm log}=\psi^{\rm log} + \gamma \psi^L$, which is precisely the same kind of ambiguity we encountered below equation \eqref{eq:lcft3} in the previous subsection. Physical quantities cannot depend on the spurious mass parameter $m_L$. Therefore, its presence in the third correlator is not at odds with conformal invariance.

We saw above that the degeneration of an operator $\psi_\eps$ with the holomorphic flux component of the stress-energy tensor $\psi^L$ suggests to regard the corresponding state $\psi^L$ as an eigenvector and the state $\psi^{\rm log}$ as a generalized eigenvector. We demonstrate this now explicitly by exhibiting the Jordan cell structure, using $L_0\psi^L=2\psi^L$ and $L_0\psi_\eps=(2+\eps)\psi_\eps$, where $L_0$ and $\bar L_0$ are the Virasoro zero mode generators.
\eq{
L_0\psi^{\rm log}=\lim_{\eps\to 0}\frac1\eps L_0(\psi_\eps-\psi^L) = \lim_{\eps\to 0}\frac{(2+\eps)\psi_\eps-2\psi_L}{\eps} = 2\psi^{\rm log} + \psi^L
}{eq:lcft14}
Similarly, $\bar L_0\psi^L=0$ and $\bar L_0\psi^{\rm log}=\psi^L$. 
Defining the Hamiltonian $H=L_0 + \bar L_0$ and the angular momentum operator $J=L_0-\bar L_0$, we see that the latter is diagonal, while the Hamiltonian acquires a Jordan cell.
\eq{
H\begin{pmatrix}
  \psi^{\rm log} \\ \psi^L
 \end{pmatrix}
= 2\,\begin{pmatrix}
   1 & 1 \\ 0 & 1
  \end{pmatrix}
 \begin{pmatrix}
  \psi^{\rm log} \\ \psi^L
 \end{pmatrix}
}{eq:lcft15}
The operators with degenerate conformal weights, $\psi^{\rm log}$ and $\psi^L$, are called ``logarithmic pair''.

Above we provided a simple example of a Jordan cell in a CFT.
More generally, Jordan cells are the defining properties of log CFTs \cite{Gurarie:1993xq} and lead to indecomposable representations of the Virasoro algebra.
As in non-hermitian quantum mechanics, also in log CFTs higher rank Jordan cells can arise. 
An important property of log CFTs is that they are necessarily non-unitary, i.e.~the Shapovalov form is not positive definite.
For review articles on log CFTs see \cite{Flohr:2001zs,Gaberdiel:2001tr}.

The kind of log CFTs we are mostly concerned with are the $c=0$ theories similar to the example above, since they arise naturally in various higher derivative gravity theories, as we shall see in section \ref{se:2}. 
The limiting procedure $\eps\to 0$ above provides a solution of the ``$c=0$ catastrophe'', see \cite{Cardy:2013rqg} and references therein. 
Log CFTs with $c=0$  were studied e.g.~in \cite{Cardy:1999zp,Gurarie:1999yx,Kogan:2001ku,Pearce:2002an,Flohr:2005dr,Mathieu:2007pe,Gaberdiel:2010rg,Gainutdinov:2012ms}, describing critical percolation and systems with quenched disorder like quenched random magnets.
We shall come back to possible condensed matter applications in section \ref{se:4.2}.

\subsection{AdS$_3$/CFT$_2$ crash course}\label{se:1.3}

The AdS/CFT correspondence \cite{Maldacena:1997re} had an important pre-cursor in three bulk dimensions \cite{Brown:1986nw}.
Brown and Henneaux showed that any 3-dimensional quantum theory of gravity with asymptotically AdS boundary conditions must be dual to a 2-dimensional CFT in the sense that the Hilbert-space falls into representations of two copies of the Virasoro algebra, whose central charges $c_{L,R}$ depend on the specific theory and the coupling constants therein. For Einstein gravity with negative cosmological constant, $\Lambda=-1/\ell^2$, the result is
\eq{
c_L = c_R = \frac{3\ell}{2G_N}
}{eq:lcft16}
where $G_N$ is Newton's constant and $\ell$ the AdS radius.
The Brown--Henneaux result \eqref{eq:lcft16} has been generalized to other 3-dimensional gravity theories with asymptotically AdS boundary conditions. 
We review now briefly and in a fairly general way how such constructions work.
For simplicity from now on we set the AdS radius to unity, $\ell=1$, and refer to ``Einstein gravity with negative cosmological constant'' simply as ``Einstein gravity''.

Step one is to identify the bulk theory.
There are no universal guidelines on what is a ``correct'' choice beyond the obvious remark that it depends on the physical situation that one would like to model.

Step two is to impose suitable boundary conditions for all the fields.
Let us focus on the metric.
A prominent example (the only one of relevance for this work) is locally asymptotically AdS boundary conditions.
Global AdS with line-element
\eq{
\extd s^2 = \extd \rho^2 -\cosh^2\!\rho\,\extd t^2 + \sinh^2\!\rho\,\extd\varphi^2
}{eq:AdS}
is topologically a solid cylinder and has an asymptotic boundary. It is useful to introduce a radial coordinate like $\rho$ on the AdS cylinder so that the asymptotic boundary is reached in the limit $\rho\to\infty$.
To reduce clutter consider Gaussian normal coordinates with respect to the radial coordinate $\rho$, like in global AdS \eqref{eq:AdS}.
Then, the condition of locally asymptotically AdS implies the asymptotic expansion of the line-element 
\eq{
\extd s^2 = \extd\rho^2 + \ga_{ij}\,\extd x^i \extd x^j\qquad \ga_{ij} = \ga_{ij}^{(0)}\,e^{2\rho} + o(e^{2\rho})_{ij} 
}{eq:lcft17}
where $x^i$ are some boundary coordinates (for instance, light-cone coordinates $x^\pm=t\pm\varphi$), $\ga_{ij}^{(0)}$ is the boundary metric and $o(e^{2\rho})$ refers to terms that diverge slower than the leading term, terms that are finite and terms that vanish in the limit $\rho\to\infty$. 
For Einstein gravity the subleading terms are restricted by the asymptotic equations of motion (EOM), which then leads to the standard Feffermann--Graham expansion where the first subleading term is finite \cite{Henningson:1998ey,deHaro:2000xn,Balasubramanian:1999re}. 
Such an expansion (after relaxing the gauge-fixing to Gaussian normal coordinates) corresponds to Brown--Henneaux boundary conditions \cite{Brown:1986nw}.
In most of the applications in the present work we are not permitted to impose Brown--Henneaux boundary conditions, but instead need subleading terms that grow polynomially in the radial coordinate $\rho$, where the power depends on the rank of the Jordan cell. We shall see this explicitly in section \ref{se:2}.

Step three is a canonical analysis analog to the one by Brown and Henneaux, see for instance \cite{Blagojevic:2002du}. 
After finding all bulk constraints and sorting them into first and second class constraints, one can construct the canonical generators of gauge transformations from suitable combinations of the first class constraints.
A key aspect of the analysis in the presence of an asymptotic boundary is that certain constraints that are first class in the bulk may fail to be first class at the boundary. 
If this happens then physical boundary states emerge.
Technically, the conversion of bulk first class constraints into second class constraints at the boundary comes from functional non-differentiability of (some of the) gauge generators.
In order to make them functionally differentiable one adds a boundary piece, namely the (variation of the) canonical boundary charges $Q$.
\eq{
\delta Q[\eps] = \oint\extd x\sqrt{|\sigma|}\, \eps\, {\cal L}(g, \pi, \delta g, \delta\pi)
}{eq:lcft18}
Here $\eps$ parametrizes the boundary condition preserving gauge transformations, $\sigma$ is the induced volume-element at the asymptotic boundary circle, and ${\cal L}$ is some function of the canonical variables (schematically denoted by $g, \pi$) and their first variations in field space.
In all applications in this paper one can drop the variation on both sides of \eqref{eq:lcft18}.
This means that the charges are integrable.\footnote{%
A baby example for an integrable charge is ${\cal L}=g\delta\pi+\pi\delta g$; similarly, a non-integrable example is ${\cal L}=g\delta\pi-\pi\delta g$. In both examples we assume state-independent $\sigma$ and $\eps$ so that $\de\sigma=\de\eps=0$.
}
One has then to check whether the charges are finite and conserved (in time).
If they fail to be integrable or finite or conserved then one has to go back to step one or two.

Step four is the derivation of the (classical) asymptotic symmetry algebra and its central charges.
The asymptotic symmetry algebra is generated by all boundary condition preserving gauge transformations modulo the trivial gauge transformations (those whose associated canonical charges are state independent).
To this end one either works out the Dirac bracket algebra of the gauge generators (including all boundary terms), which is straightforward but somewhat tedious, or one exploits the relation $\{Q[\eps_1], \,Q[\eps_2]\}=\delta_{\eps_2} Q[\eps_1]$ and calculates the gauge variation of the canonical charges \cite{Brown:1986ed,Brown:1986nw}.
A standard result for one chiral half of the charges is $\de_\eps{\cal L}=2\eps{\cal L}'+\eps^\prime{\cal L}+\frac{k}{\pi}\,\eps'''$, with some constant $k$ that depends on the coupling constants of the theory. Making a Fourier-decomposition ${\cal L}=\sum_n L_n e^{-inx^+}$ then leads to the Virasoro algebra
\eq{
[L_n,\,L_m]=(n-m)\,L_{n+m}+\frac{c_L}{12}\,(n^3-n)\,\de_{n+m,\,0}
}{eq:lcft19}
with central charge $c_L=6k$.
A similar result is obtained for the other chiral half, which leads to a Virasoro algebra with central charge $c_R$.
In some cases of AdS holography the asymptotic symmetry algebra can be larger than two copies of the Virasoro algebra. 
In this review we focus mostly on the simpler cases where the asymptotic symmetry algebra is just the conformal algebra.

Step five improves the classical asymptotic symmetry algebra to a quantum asymptotic symmetry algebra in order to maintain consistency with the Jacobi identities beyond the (semi-)classical approximation.
This step is not necessary in theories like Einstein gravity or TMG.
Simple examples where it is necessary are spin-3 gravity \cite{Henneaux:2010xg,Campoleoni:2010zq} and conformal Chern--Simons gravity \cite{Afshar:2011yh,Afshar:2011qw}; see \cite{Afshar:2012nk,Afshar:2012hc} for more complicated examples in higher spin gravity.

After completing these steps the dual CFT can be constrained.
The main information available at this stage is the symmetry algebra of the dual CFT (two copies of Virasoro and possibly more) and the central charges.
There are two interesting special cases.
If $c_L-c_R=0$ then there is no diffeomorphism anomaly.
This happens for example in Einstein gravity.
If $c_L+c_R=0$ then there is no trace anomaly.
This happens for example in conformal Chern--Simons gravity.

A simple check of the AdS$_3$/CFT$_2$ correspondence is the calculation of 2- and 3-point correlators of the stress-energy tensor on the gravity side and to relate them to corresponding CFT correlators using the standard AdS/CFT dictionary \cite{Witten:1998qj}. For example,
\eq{
\langle \psi^{L/R}\, \psi^{L/R}\rangle_{\textrm{\tiny CFT}} \sim \frac{\de^2\Gamma_{\textrm{\tiny grav}}}{\de\psi^{L/R}_{\textrm{\tiny NN}}\de\psi^{L/R}_{\textrm{\tiny NN}}}
}{eq:lcft20}
where the left hand side contains the 2-point correlator between two (anti-)holomorphic flux components of the stress-energy tensor, $\psi^{L/R}$, and the right hand side the second variation of the holographically renormalized on-shell action $\Gamma_{\textrm{\tiny grav}}$ (including all boundary terms) with respect to non-normalizable left-(right-)moving solutions $\psi^{L/R}_{\textrm{\tiny NN}}$ of the linearized EOM on the AdS background \eqref{eq:AdS}.
From the CFT perspective these correlators are constrained (essentially uniquely) by the conformal Ward identities and the values of the central charges. If one believes in the correspondence there is actually nothing to calculate.
Historically, it was nevertheless useful to calculate correlators like \eqref{eq:lcftc1}-\eqref{eq:lcftc3} to exhibit explicitly the log CFT structure on the gravity side \cite{Skenderis:2009nt,Skenderis:2009kd,Grumiller:2009mw,Grumiller:2009sn,Alishahiha:2010bw}, thereby providing evidence for the AdS$_3$/log CFT$_2$ correspondence discussed in section \ref{se:2}.
We shall be more explicit about gravity correlators, the linearized gravity modes $\psi^{L/R}_{\textrm{\tiny NN}}$ and their normalizable counterparts in that section.

Another check of the AdS$_3$/CFT$_2$ correspondence is to compare partition functions.
If the correspondence holds then the Euclidean CFT partition function on the torus must be identical with the Euclidean quantum gravity partition function evaluated on the filled AdS torus (up to overall normalization).
\eq{
Z_{\textrm{\tiny CFT}}(q,\, \bar q) \sim Z_{\textrm{\tiny grav}}(q, \,\bar q)
}{eq:lcft21} 
Here $q=e^{i\tau}$ is the modular parameter. On the gravity side $\tau=\theta + i \beta$, where $\theta$ is the angular potential and $\beta$ the inverse temperature (or periodicity in Euclidean time). 
As a first step, it is already of interest to calculate the classical and 1-loop partition functions on the gravity side.
For Einstein gravity this works as follows.
If the semi-classical approximation holds, the partition function can be expanded perturbatively,
\eq{
Z_{\textrm{\tiny grav}}(q, \,\bar q) = e^{-k\Ga^{(0)}+\Ga^{(1)}+\frac1k\,\Ga^{(2)}+\dots}
}{eq:lcft22}
where $k$ is essentially the inverse Newton constant, which becomes a large parameter semi-classically.
Thus, the partition function is dominated by the classical saddle-points captured by the classical on-shell action $\Ga^{(0)}$. The Euclidean AdS saddle-point in Einstein gravity leads to a classical partition function given by (see e.g.~\cite{Maldacena:1998bw})
\eq{
e^{-k\Ga^{(0)}} = (q\bar q)^{-\frac k4}\,.
}{eq:lcft23}
The 1-loop contribution can be calculated efficiently using heat-kernel methods \cite{Vassilevich:2003xt}.
The result is (see e.g.~\cite{Giombi:2008vd,David:2009xg}; for earlier papers see \cite{Camporesi:1990wm,Camporesi:1994ga,Mann:1996ze,Bytsenko:1994bc})
\eq{
e^{\Ga^{(1)}} = \prod_{n=2}^\infty\frac{1}{(1-q^n)(1-\bar q^n)}\,.
}{eq:lcft24}
Neglecting higher loops (or arguing that they vanish, see for instance \cite{Maloney:2007ud}) the quantum gravity partition function is then given by
\eq{
Z_{\textrm{\tiny grav}}(q, \,\bar q) = |q|^{-\frac k2}\,\prod_{n=2}^\infty\frac{1}{|1-q^n|^2}
}{eq:lcft25}
By comparison, the CFT partition function that counts the Virasoro descendants of the vacuum is given by the Virasoro vacuum character (see e.g.~\cite{diFrancesco})
\eq{
Z_{\textrm{\tiny CFT}}(q, \,\bar q) = \tr\big(q^{L_0}\bar q^{\bar L_0}\big)= |q|^{-\frac{c}{12}}\,\prod_{n=2}^\infty\frac{1}{|1-q^n|^2}
}{eq:lcft26}
provided that the ground state has conformal weights $h=\bar h = -\frac{c}{24}=-\frac k4$. (The Brown--Henneaux analysis above yielded $c=6k$.)
Therefore, in Einstein gravity at 1-loop level the quantum gravity partition function that counts all perturbative excitations above the AdS ground state coincides precisely with the corresponding CFT partition function.

The checks above, while non-trivial, are still fairly basic checks of the AdS$_3$/CFT$_2$ correspondence and by no means sufficient for its validity.
Moreover, they are generic checks, in the sense that apart from the values of the central charges and possible further contributions to the partition function they do not depend on the specific CFT that is supposed to be dual to the gravitational theory.
The advantage of such generic checks is that any purported AdS$_3$/CFT$_2$ correspondence can be subjected to them and the calculations have fairly universal applicability.
The disadvantage is that no specific features of a suggested correspondence are tested.
The situation changes once a concrete proposal for a CFT is available.
An example for such a proposal is the one by Witten \cite{Witten:2007kt}, which would have related Einstein gravity at certain values of the inverse Newton constant $k$ to so-called extremal CFTs, like the monster theory by Frenkel, Lepowsky and Meurman.
While this proposal did not work, see for instance \cite{Maloney:2007ud,Gaberdiel:2007ve,Gaberdiel:2008pr}, it inspired the community to think about possible resolutions of the problems encountered with Einstein gravity.
One potential solution is to consider more general theories of gravity than Einstein gravity.
This is the subject of section \ref{se:2}.

\section{AdS$_3$/log CFT$_2$ correspondence}\label{se:2}

In this section we focus on the AdS$_3$/log CFT$_2$ correspondence.
The first papers to suggest an AdS$_3$/log CFT$_2$ correspondence used a higher derivative action for a scalar field on an AdS background \cite{Ghezelbash:1998rj} that was identified as a singleton dipole by Kogan \cite{Kogan:1999bn}, see also \cite{Myung:1999nd,Lewis:1999qv}.
These log CFTs are not of the type discussed in section \ref{se:1.2}, because the stress-energy tensor can only acquire a logarithmic partner if there are at least two spin-2 modes with degenerate weights.
In higher derivative gravity theories additional spin-2 modes as compared to Einstein gravity arise automatically. As we shall review in this section the degeneration required for a log CFT is then possible \cite{Grumiller:2008qz}.

In subsection \ref{se:2.1} we recall the critical TMG story that motivated the correspondence and show how a Jordan cell arises in the linearized excitations on the gravity side.
In subsection \ref{se:2.2} we mention various checks of the correspondence, mostly by using well-known properties of log CFTs as input and trying to recover them on the gravity side. 
We also comment on chiral gravity.
In subsection \ref{se:2.3} we go beyond TMG and show that fairly generic 3-dimensional higher derivative theories of gravity exhibit critical points that lead to Jordan cells in the graviton sector.

\subsection{Critical TMG --- chiral or log?}\label{se:2.1}

The TMG action \cite{Deser:1982vy,Deser:1982wh,Deser:1982a}
\begin{multline}
\Gamma_{\textrm{\tiny TMG}} = \frac{1}{16\pi G_N}\,\int_M \extd^3x \Big[\sqrt{-g}\,\Big(R+\frac{2}{\ell^2}\Big) + \frac{1}{2\mu}\,\epsilon^{\la\mu\nu}\,\Ga^\si{}_{\la\rho}\,\Big(\partial_\mu\Ga^\rho{}_{\nu\si}+\frac23\,\Ga^\rho{}_{\mu\tau}\Ga^\tau{}_{\nu\si}\Big)\Big] \\
+ \frac{1}{8\pi G_N}\,\int_{\partial M}\!\!\!\extd^2x\sqrt{-\ga}\,\Big[K - \frac{1}{\ell} + \frac{1}{4\mu}\,\Big(K^{\al\be} K_{\al\be}-\frac12\,K^2\Big)\Big] 
\label{eq:lcft27}
\end{multline}
has two dimensionless combinations of coupling constants, $\ell/G_N$ and $\mu\ell$.
The quantity $\mu$ is the Chern--Simons coupling constant.
The three boundary terms in the second line are Gibbons--Hawking--York, holographic cosmological constant and Guica--Skenderis--Taylor--van Rees counterterm \cite{Guica:2010sw}, respectively.
TMG has one local physical degree of freedom, the massive graviton, for any non-singular value of the coupling constants.
For generic values of the coupling constants it turns out that BTZ black holes \cite{Banados:1992wn,Banados:1992gq} have positive energy but the massive gravitons negative energy \cite{Li:2008dq}. (The original work by Deser, Jackiw and Templeton considered the opposite overall sign in the action \eqref{eq:lcft27} and set the cosmological constant to zero, $1/\ell=0$, so that massive gravitons had positive energy.)

Performing the Brown--Henneaux-type of analysis outlined in section \ref{se:1.3} for TMG leads to two copies of the Virasoro algebra \eqref{eq:lcft19} with central charges \cite{Kraus:2005zm}
\eq{
c_L = \frac{3\ell}{2G_N}\,\big(1-\frac{1}{\mu\ell}\big)\qquad c_R = \frac{3\ell}{2G_N}\,\big(1+\frac{1}{\mu\ell}\big)\,.
}{eq:lcft28}
Critical TMG (cTMG) is obtained for the tuning
\eq{
\mu\ell = 1\,.
}{eq:lcft29}
In that case the left central charge vanishes, $c_L=0$.
If the dual CFT exists and is unitary then it must be chiral.

Motivated by the failure of Einstein gravity to produce a partition function that factorizes holomorphically \cite{Maloney:2007ud}, Li, Song and Strominger \cite{Li:2008dq} proposed to consider chiral gravity, which is cTMG with Brown--Henneaux boundary conditions \cite{Maloney:2009ck}, and was conjectured to be dual to a chiral CFT. 
If chiral gravity exists it is a simple non-trivial quantum theory of gravity.
For the smallest possible value of the central charge, $c_R=24$, its partition function is given by the $J$-function \cite{Witten:2007kt,Maloney:2009ck}
\eq{
Z_{\chi \textrm{G}}(\bar q)=J(\bar q)=\frac{1}{\bar q} + 196884 \,\bar q + {\cal O}(\bar q^2)\,.
}{eq:chiral}
The number $196884$ is interpreted as one Virasoro descendant of the vacuum plus $196883$ primary fields corresponding to BTZ black hole microstates \cite{Witten:2007kt}. 
The chiral gravity quantum entropy $S_{\chi\textrm{G}}=\ln{196883}\approx 12.2$ differs only by about $3\%$ from the semi-classical Bekenstein--Hawking result $S_{\textrm{\tiny BH}}=4\pi\approx 12.6$.
Since $c_R=24$ is a fairly sizable value for the central charge the semi-classical approximation is expected to be good.
For quantum gravity applications the chiral gravity situation seems optimal: there are quantum corrections that are not completely negligible (of the order of a few percent), but the theory is not ``ultra-quantum'' so that geometric notions associated with the semi-classical limit, like black hole horizons, could still be discussed meaningfully.

One consequence of the chiral gravity conjecture was the absence of massive graviton excitations, a result which was disputed by Carlip, Deser, Waldron and Wise \cite{Carlip:2008jk} who found local excitations even at the critical point. This apparent contradiction engendered an intense discussion \cite{Grumiller:2008qz,Li:2008yz,Carlip:2008eq,Giribet:2008bw,Strominger:2008dp,Grumiller:2008es,Henneaux:2009pw,Maloney:2009ck}.

Before addressing the resolution we need a few technical details and checks.
We shall come back to chiral gravity at the end of subsection \ref{se:2.2}.
The linearized EOM in transverse gauge, $\nabla_\mu (\psi^{\mu\nu}-g^{\mu\nu}\psi^\la{}_\la)=0$, for the graviton excitations $\psi_{\mu\nu}$ around the AdS background \eqref{eq:AdS} can be written as \cite{Li:2008dq}
\eq{
\big({\cal D}^L{\cal D}^R{\cal D}^{(\mu)}\psi\big)_{\mu\nu} = 0
}{eq:lcft30}
with the mutually commuting first order operators ${\cal D}^{L/R}={\cal D}^{(\mu)}|_{\mu=\pm1/\ell}$ and
\eq{
 ({\cal D}^{(\mu)})_\al{}^\be=\eps_\al{}^{\ga\be}\,\nabla_\ga + \mu\,\de_\al^\be\,.
}{eq:lcft31}
The linearized EOM \eqref{eq:lcft30} imply tracelessness of all linearized solutions in transverse gauge, which is why this gauge choice is also referred to as ``transverse-traceless''.
Modes $\psi^M_{\al\be}$ annihilated by operators like ${\cal D}^{(\mu)}$ can have different properties, like regularity or normalizability.
In light-cone coordinates $x^\pm=t\pm\varphi$ the separation ansatz
\eq{
\psi^M_{\al\be} = e^{-ihx^+ - i\bar hx^-}\,F_{\al\be}(\rho)
}{eq:lcft32}
defines the $sl(2)$ weights $h$ and $\bar h$.
Namely, the Virasoro zero mode generators are two of the six AdS Killing vectors, $L_0=i\partial_+$ and $\bar L_0=i\partial_-$, so that $L_0\psi^M=h\psi^M$ and $\bar L_0\psi^M=\bar h\psi^M$.
The remaining four Killing vectors span the remaining $sl(2)$ generators $L_{\pm 1}$ and $\bar L_{\pm 1}$, see e.g.~\cite{Li:2008dq} for explicit expressions.
For each value of $\mu$ there is a unique mode $\psi^M$ that is regular, normalizable and obeys the $sl(2)$ primary conditions $L_1\psi^M=\bar L_1\psi^M=0$ \cite{Li:2008dq}. 
For positive $\mu$ the $sl(2)$ weights of these primaries are given by
\eq{
\big(h, \,\bar h\big) = \big(\frac32+\frac{\mu\ell}{2}, \,-\frac12+\frac{\mu\ell}{2}\big)\,.
}{eq:lcft33}
The difference of the weights, $|h-\bar h|=2$, always is two, as it must be for a graviton excitation.
For negative $\mu$ one obtains instead $\big(h, \,\bar h\big) = \big(-\frac12-\frac{\mu\ell}{2}, \,\frac32-\frac{\mu\ell}{2}\big)$.
This means the primaries $\psi^L$ [$\psi^R$] annihilated by ${\cal D}^L$ [${\cal D}^R$] have weights $(2, 0)$ [$(0, 2)$] and can be interpreted on the CFT side as $L_{-2}$ [$\bar L_{-2}$] descendants of the vacuum.
For the critical tuning \eqref{eq:lcft29} the operators ${\cal D}^L$ and ${\cal D}^{(\mu)}$ degenerate with each other.
The same is true for the weights of the primaries $\psi^L$ and $\psi^M$.

A key observation is that the degeneracy of the left-moving boundary graviton $\psi^L$ with the massive graviton $\psi^M$ implies the existence of a log mode  \cite{Grumiller:2008qz}
\eq{
\big({\cal D}^L{\cal D}^L\psi^{\rm log}\big)_{\mu\nu} = 0 \qquad \big({\cal D}^L\psi^{\rm log}\big)_{\mu\nu} \propto \psi^L_{\mu\nu}\,.
}{eq:lcft34}
We have thus precisely the same features as for the degeneracy of two eigenvectors, where a generalized eigenvector emerges \eqref{eq:lcft3}.
Given the general analysis of section \ref{se:1} one should therefore expect the appearance of a Jordan cell in cTMG.
In fact, using a construction analog to \eqref{eq:lcft13} (with $\eps=\frac{\mu\ell-1}{2}$)\footnote{\label{fn:4}
The mode presented in \eqref{eq:lcft35} differs by a factor $2$ from the mode in the original paper \cite{Grumiller:2008qz} and some of the early literature.
We choose the current normalization since it yields correlators with the same numerical coefficients as in \eqref{eq:lcftc2}, \eqref{eq:lcftc3}.
Note that the prefactor $y=-2(it+\ln\cosh\rho)$ on the right hand side of \eqref{eq:lcft35} obeys the primary conditions $L_1y=\bar L_1 y=0$ and has $sl(2)$ weights $(1,\, 1)$.
}
\eq{
\psi^{\rm log}_{\al\be} = \lim_{\eps\to 0}\frac{\psi^M_{\al\be}-\psi^L_{\al\be}}{\eps} = -2(it+\ln\cosh\rho)\,\psi^L_{\al\be}
}{eq:lcft35}
yields the Jordan cell \eqref{eq:lcft15} for the Hamiltonian $H=L_0+\bar L_0=i\partial_t$ and a diagonalizable matrix for the angular momentum operator $J=L_0-\bar L_0=i\partial_\varphi$.
This observation provided the basis for the conjecture that cTMG is dual to a log CFT with $c_L=0$ and $c_R=3\ell/G_N$ \cite{Grumiller:2008qz}, where the left-moving part of the stress-energy tensor acquires a logarithmic partner.

\subsection{Checks and features of the correspondence}\label{se:2.2}

The first basic checks of the AdS/log CFT correspondence for cTMG were performed in \cite{Grumiller:2008qz}. 
A potentially worrying feature of the log mode \eqref{eq:lcft35} is its linear dependence on time and its asymptotic growth proportional to $\rho$ as compared to its partner $\psi^L$.
On general grounds, this might lead to infinite or time-dependent energy of these linearized modes.
However, it turns out that both the time-dependence and the $\rho$-divergence drop out, and the energy associated with the log modes is finite, time-independent and negative.
The last property was interpreted as a sign of non-unitarity, as expected from the log CFT side.
(A comparable precedent regarding the time-(in-)dependence of the energy is the static spherically symmetric solution of the Einstein-massless-Klein-Gordon model with a scalar field that grows linearly in time \cite{Wyman:1981bd}.)

A related worry is the possible divergence of the holographic stress-energy tensor.
Using a Fefferman--Graham type of expansion \eqref{eq:lcft17} with \cite{Grumiller:2008qz,Grumiller:2008es} (we set $\ell=1$)
\eq{
\ga_{ij} = \ga_{ij}^{(0)}\,e^{2\rho} + \ga_{ij}^{(1)}\,\rho + \ga_{ij}^{(2)} + \dots 
}{eq:lcft36}
and the chirality condition 
\eq{
\ga_{ij}^{(1)}-\ga_{ik}^{(1)}\ga^{kn}_{(0)}\epsilon_{ni}=0 
}{eq:chirality}
as well as tracelessness, $\ga^{(2)}_{ij}\ga_{(0)}^{ij}=0$, it can be shown that all terms linear in $\rho$ or $t$ drop out from the holographic stress-energy tensor \cite{Grumiller:2008qz,Ertl:2009ch}.
\eq{
T_{ij} \propto \ga_{ij}^{(2)} - \ga_{ik}^{(2)}\ga^{kn}_{(0)}\epsilon_{nj}  + \ga_{ij}^{(1)} + (i\leftrightarrow j)
}{eq:lcft37}
The overall normalization depends on conventions; in the conventions of \cite{Kraus:2005zm} it is $1/(16\pi G_N)$.
The stress-energy tensor \eqref{eq:lcft37} is traceless, finite and conserved \cite{Grumiller:2008qz} but not chiral \cite{Henneaux:2009pw,Maloney:2009ck,Skenderis:2009nt,Ertl:2009ch}.
(If the boundary metric is curved then $\ga_{ij}^{(2)}$ no longer is traceless and the stress-energy tensor acquires an additional contribution \cite{Henningson:1998ey,deHaro:2000xn,Balasubramanian:1999re,Kraus:2005zm,Skenderis:2009nt}.)
The non-chirality can be seen directly from the form of the stress-energy tensor \eqref{eq:lcft37}: the first two terms project out any left-moving part, so together they contribute only to the right-moving part of the stress-energy tensor; the last term $\ga_{ij}^{(1)}$ contributes only to the left-moving part of the stress-energy tensor due to the chirality condition \eqref{eq:chirality}.
The boundary conditions proposed in \cite{Grumiller:2008qz,Grumiller:2008es} also lead to a well-defined variational principle, in the sense that the first variation of the action vanishes for all variations that preserve the boundary conditions.

Another potential caveat is that the log modes \eqref{eq:lcft35} might not be there in the first place, either because they are pure gauge or because they are artifacts of linearization.
However, for any vector field $\xi^\al$ that preserves the transverse gauge condition the quantity $\nabla_{(\al}\xi_{\be)}$ solves the linearized Einstein equations, but due to \eqref{eq:lcft34} the log modes do not solve the linearized Einstein equations. Therefore, the log modes cannot be pure gauge. Instead, they correspond to the ``massive graviton'' of TMG at the critical point, concurrent with \cite{Carlip:2008jk}. 
The presence of precisely one local physical degree of freedom in cTMG beyond linearization was shown in \cite{Grumiller:2008pr} by a canonical analysis. 
Carlip generalized this analysis and included boundary terms \cite{Carlip:2008qh}, which was important for the considerations in \cite{Henneaux:2009pw,Maloney:2009ck}, in particular for the canonical boundary charges.
Exact solutions of cTMG with the boundary behavior of log modes \eqref{eq:lcft35} were constructed in \cite{Garbarz:2008qn}, see also \cite{Chow:2009km,Giribet:2011vv,Sachs:2011xa}.

Quasi-normal modes provide another valuable AdS/CFT tool. 
Birmingham, Sachs and Solodukhin showed exact agreement between the quasi-normal frequencies of linear perturbations in a BTZ background and the momentum space poles of the retarded propagator of the corresponding operators in the dual CFT \cite{Birmingham:2001pj} (see \cite{Sachs:2008gt} for an application to TMG).
In a log CFT where two operators degenerate with each other, on general grounds one might expect the emergence of double poles in the retarded 2-point correlator of two log modes.
Sachs has shown that this indeed happens \cite{Sachs:2008yi}, and that these double poles are responsible
for the linear time dependence [as in \eqref{eq:lcft35}] of the corresponding quasi-normal modes.

After these more or less basic checks the strongest outstanding test of the AdS/log CFT correspondence was the verification of the conformal Ward identities on the gravity side. 
Indeed, 2-point correlators on the gravity side match precisely the log CFT results \eqref{eq:lcftc1}-\eqref{eq:lcftc3}, as shown by Skenderis, Taylor and van Rees \cite{Skenderis:2009nt} and confirmed independently in \cite{Grumiller:2009mw} by calculating the non-normalizable modes $\psi^{L/R/\textrm{log}}_{\textrm{\tiny NN}}$ and evaluating the right-hand side of correlators like \eqref{eq:lcft20}. 
The latter reference also calculated 3-point correlators and again found a precise matching with 3-point correlators in a log CFT \cite{Kogan:2001ku}.
The result for the new anomaly appearing in these correlators is \cite{Skenderis:2009nt,Grumiller:2009mw}
\eq{
b_L = -\frac{3\ell}{G_N}\,.
}{eq:lcft38}

Another interesting check is the comparison of partition functions analog to \eqref{eq:lcft21}.
This was done in \cite{Gaberdiel:2010xv}.
Unfortunately, the log CFT partition function is not sufficiently known. 
In particular, it is not clear how to systematically take into account multi-log excitations since the corresponding operator is a non-chiral logarithmic field.
Nevertheless, one can compare the partition functions that take into account all the Virasoro descendants of the  low-lying states, namely the Virasoro descendants of the vacuum and of the log operator, neglecting all the  multi-log excitations.
The result on the log CFT side is 
\eq{
Z_{\textrm{\tiny CFT}}^{(1)}(q, \,\bar q) = \prod_{n=2}^\infty\frac{1}{|1-q^n|^2}\, \big(1+\frac{q^2}{|1-q|^2}\big) + \dots
}{eq:lcft39}
where the superscript and the ellipsis indicate that we took into account only single log excitations and also neglected the overall factor $q^{-\tfrac{c}{12}}$ so that we can directly compare with the 1-loop partition function on the gravity side
\eq{
Z_{\textrm{\tiny cTMG}}^{(1)}(q, \,\bar q) = \prod_{n=2}^\infty\frac{1}{|1-q^n|^2}\, \big(1+\frac{q^2}{|1-q|^2}\big) + \sum_{h,\,\bar h} N_{h,\,\bar h}q^h\bar q^{\bar h}\,\prod_{n=1}^\infty\frac{1}{|1-q^n|^2}\,.
}{eq:lcft40}
The last term is presented in a way that makes it evident that it describes the character of the $(h, \bar h)$ representation of the Virasoro algebra, where $N_{h,\,\bar h}$ is the multiplicity with which this representation occurs. 
Setting to zero all multiplicity coefficients $N_{h,\,\bar h}$ shows perfect agreement between the partition functions \eqref{eq:lcft39} and \eqref{eq:lcft40}.
Thus, we have precise agreement of partition functions between log CFT and cTMG for single-log excitations.
If our interpretation is correct then the multiplicity coefficients $N_{h,\,\bar h}$ should account for the multi-log excitations on the gravity side.
For consistency all these numbers then must be non-negative. 
Indeed, it can be shown that they are \cite{Gaberdiel:2010xv}.
The 1-loop calculation in cTMG thus leads to a result that is perfectly consistent with the proposal that the dual CFT is logarithmic and provides further support of its validity.

Before ending this subsection we come back to chiral gravity \cite{Li:2008dq}.
Given that the log modes \eqref{eq:lcft35} require the weaker fall-off behavior \eqref{eq:lcft36} as compared to Brown--Henneaux, it was suggested to impose Brown--Henneaux boundary conditions to define chiral gravity, a truncated version of cTMG where the log modes are absent \cite{Grumiller:2008qz,Li:2008yz,Strominger:2008dp,Maloney:2009ck}.
In this sense, the results by Li, Song and Strominger \cite{Li:2008dq} are not at odds with the results by Carlip, Deser, Waldron and Wise \cite{Carlip:2008jk}, since the massive graviton modes exist as log modes \cite{Grumiller:2008qz} even for cTMG, but they are absent if eliminated by stricter boundary conditions or some other means.
Interestingly, Giribet, Kleban and Porrati were able to construct a linearized log mode (a descendant of the ``log-primary'') that was compatible with Brown--Henneaux boundary conditions \cite{Giribet:2008bw}, which at that time appeared to be at odds with the chiral gravity conjecture.
This issue was clarified by Maloney, Song and Strominger, who showed that there is a linearization instability \cite{Maloney:2009ck}, so that imposing Brown--Henneaux boundary conditions seems still good enough to eliminate the log modes.
Thus, chiral gravity can exist as a left-charge-equal-to-zero superselection sector of cTMG.\footnote{
On the gravity side there is the following caveat. Compere, de~Buyl and Detournay found exact solutions of chiral gravity that do not solve the Einstein equations \cite{Compere:2010xu}.
There is some evidence that all these solutions are singular, either due to curvature or causal singularities (closed time-like curves), in which case they would not be relevant in the dual CFT.
However, if there were regular solutions this would be problematic for chiral gravity, which takes into account only Einstein solutions \cite{Maloney:2009ck}. 
}
For $c_R=24$ it was suggested that chiral gravity is dual to one chiral half of the monster CFT  with partition function \eqref{eq:chiral} \cite{Li:2008dq,Maloney:2009ck}.
Since chiral gravity requires the critical tuning \eqref{eq:lcft29} it is of interest to check whether or not such a tuning is stable under RG flow.
Percacci and Sezgin calculated the 1-loop $\beta$-functions in TMG and showed that the critical tuning \eqref{eq:lcft29} is not stable in general \cite{Percacci:2010yk}.
They also showed that the product of Chern--Simons coupling and Newton constant, $\mu G_N$, has vanishing $\beta$-function.
Therefore, a scaling limit of TMG, $G_N\to\infty$ keeping finite the product $\mu G_N$, should be stable under RG flow.
This motivated the holographic investigation of conformal Chern--Simons gravity \cite{Afshar:2011yh,Afshar:2011qw}.
Recently, it was suggested \cite{Bagchi:2012yk} that flat-space boundary conditions in conformal Chern--Simons gravity could lead to a dual CFT with the same properties as the conjectured dual CFT of chiral gravity.
It remains to be seen whether (flat-space) chiral gravity passes further tests.

\subsection{Jordan cells in other higher-derivative gravity theories}\label{se:2.3}

This subsection presents other models of critical gravity that have been investigated in the last years. 
We briefly review these models, point out similarities and differences to the rank-2 case of cTMG, and refer the reader to further literature where checks --- in line with the previous subsection --- have been performed.

\subsubsection{New massive gravity}

The first parity even candidate theory with critical behavior was new massive gravity (NMG) \cite{Bergshoeff:2009hq,Bergshoeff:2009aq}.
The bulk action is given by
\begin{equation}
 S_{\rm NMG} = \frac{1}{16\pi G_N}\int{\rm d}^3x \,\sqrt{-g}\,\Big[\sigma R -2\lambda m^2 +
          \frac{1}{m^2}\Big(R^{\mu\nu}R_{\mu\nu}-\frac38 R^2\Big)\Big]\,.
\end{equation}
NMG propagates two massive helicity-2 modes around an AdS background and has two massless
Einstein modes, which as usual only contribute to boundary graviton excitations, but not to bulk degrees of freedom. The AdS radius is given by $1/\ell^2=2m^2(\sigma\pm\sqrt{1+\lambda})$. 
In transverse-traceless gauge the linearized EOM take the form
\begin{equation}
 (\mathcal{D}^L\mathcal{D}^R\mathcal{D}^{m_+}\mathcal{D}^{m_-}\psi)_{\mu\nu}=0 
\end{equation}
where $\mathcal{D}^{L/R}$ are exactly as in \eqref{eq:lcft30} and $\mathcal{D}^{m_\pm}$ as in \eqref{eq:lcft31} with $\mu=m_\pm$.
The mass parameters\footnote{%
The parameters $m_\pm$ do not denote the physical mass. 
The physical mass is given by $m_{\rm phys}^2=m_\pm^2-1/\ell^2$, while the sign of $m_\pm$ is related to the mode's helicity.
}
$m_\pm$ are related to the parameter $m^2$ and the AdS radius as $m_\pm\ell=\pm\sqrt{1/2-\sigma m^2\ell^2}$.
At its critical point, $\sigma m^2\ell^2=-1/2$, the two massive operators $\mathcal{D}^{m_\pm}$ degenerate
with $\mathcal{D}^L$ and $\mathcal{D}^R$ respectively. Critical NMG is very similar to cTMG but now both
the holomorphic and the anti-holomorphic flux components of the stress-energy tensor acquire logarithmic partners to form rank-2 Jordan cells \eqref{eq:lcft15}. Many checks and calculations that have been performed for cTMG have been done for critical NMG as well, see \cite{Liu:2009bk,Liu:2009kc,AyonBeato:2009yq,Clement:2009ka,Grumiller:2009sn,
Hohm:2010jc,Alishahiha:2010bw,Giribet:2010ed,Gaberdiel:2010xv,Myung:2011bn,Sen:2012fc,Cunliff:2013en}.

\subsubsection{Generalized massive gravity}

Combining TMG and NMG leads to generalized massive gravity (GMG) \cite{Bergshoeff:2009hq,Bergshoeff:2009aq}
(for earlier work on linearized GMG see \cite{Deser:2002iw}).
Its bulk action is given by
\begin{multline}
 S_{\rm GMG} = \frac{1}{16\pi G_N}\int{\rm d}^3x \,\sqrt{-g}\,\Big[\sigma R -2\lambda m^2 +
          \frac{1}{m^2}\Big(R^{\mu\nu}R_{\mu\nu}-\frac38 R^2\Big) \\
+ \frac{1}{2\mu}\,\epsilon^{\la\mu\nu}\,\Ga^\si{}_{\la\rho}\,\Big(\partial_\mu\Ga^\rho{}_{\nu\si}+\frac23\,\Ga^\rho{}_{\mu\tau}\Ga^\tau{}_{\nu\si}\Big)\Big]\,.
\end{multline}
In the limit $m^2\to\infty$ ($\mu\to\infty$) TMG (NMG) is recovered.
The scaling limit $m\to 0$, keeping fixed $m^2 G_N$, leads to pure fourth-order gravity \cite{Deser:2009hb};
the scaling limit $\mu\to 0$, keeping fixed $\mu G_N$, leads to conformal Chern--Simons gravity, see \cite{Afshar:2011yh,Afshar:2011qw} and references therein.
Below we assume that none of these special limits is taken.

Like NMG, GMG has two Einstein modes and two massive ones. 
Unlike NMG, the two massive modes are not related by parity.
The EOM allow again an AdS background with AdS radius $1/\ell^2=2m^2(\sigma\pm\sqrt{1+\lambda})$.
The linearized EOM take the form
\begin{equation}
 (\mathcal{D}^L\mathcal{D}^R\mathcal{D}^{m_1}\mathcal{D}^{m_2} \psi)_{\mu\nu}=0 
\end{equation}
where again $\mathcal{D}^{L/R}$ are exactly as in \eqref{eq:lcft30} and $\mathcal{D}^{m_{1,2}}$ as in \eqref{eq:lcft31} with $\mu=m_{1,2}$.
The mass parameters $m_{1,2}$ are given by
\begin{equation}
 m_{1,2}\ell=\frac{m^2\ell^2}{2\mu\ell}\pm\sqrt{\frac12-\sigma m^2\ell^2+\frac{m^4\ell^4}{4\mu^2\ell^2}} \,.
\end{equation}
The possibility to tune the masses $m_1$ and $m_2$ independently leads to a richer structure than in NMG and more degenerations and critical points are available \cite{Liu:2009pha,Grumiller:2010tj,Bergshoeff:2010iy}.

Critical theories arise if both masses take the same value or if one or both of them obeys $m_i\ell=\pm1$. In
addition to the log CFTs that were discussed up to now we have the possibilities, 1.~to have two massive modes degenerate if $\mathcal{D}^{m_1}=\mathcal{D}^{m_2}$ ($m_1=m_2\neq\pm1/\ell$), or 2.~to have three modes degenerate $\mathcal{D}^{m_1}=\mathcal{D}^{m_2}=\mathcal{D}^{L/R}$ ($m_1=m_2=\pm1/\ell$).
In the first case we obtain a gravity dual to a log CFT with non-vanishing central charges.
The second case presents a gravity dual to a rank-3 log CFT. 

\begin{figure}
\begin{center}
\setlength{\unitlength}{1mm}
\begin{picture}(110,110)(-10,-10)
\thicklines
\put(40,-10){\vector(0,1){100}} 
\put(40,95){\makebox(0,0){$m_2\ell$}}

\put(-10,40){\vector(1,0){100}} 
\put(95,40){\makebox(0,0){$m_1\ell$}}

\dashline{2}(-10,50)(85,50)
\put(85,53){\makebox(0,0){$c_L =0 $}}

\dashline{2}(50,-10)(50,85)
\put(55,90){\makebox(0,0){$c_L =0 $}}

\dashline{2}(-10,30)(85,30)
\put(25,90){\makebox(0,0){$c_R =0 $}}

\dashline{2}(30,-10)(30,85)
\put(85,27){\makebox(0,0){$c_R =0 $}}

\dashline{1}(0, 0)(75,75)
\put(80,80){\makebox(0,0){$m_1 = m_2$}}

\dottedline{2}(0, 80)(75 ,5)
\put(80,0){\makebox(0,0){NMG}}

\put(40,40){\circle*{3}}
\put(50,50){\circle{3}}
\put(50,50){\circle*{2}}

\put(30,30){\circle{3}}
\put(30,30){\circle*{2}}
 
\qbezier(42,15)(42,38)(65,38)
\qbezier(38,65)(38,42)(15,42)
\dashline{5}(65,38)(85, 38.5)
\dashline{5}(42,15)(41.5, -5)
\dashline{5}(38,65)(38.5, 85)
\dashline{5}(15,42)(-5, 41.5)
\end{picture}
\caption{Special loci in the GMG parameter space. See main text for explanations. }
 \label{fig:2}
\end{center}
\end{figure}
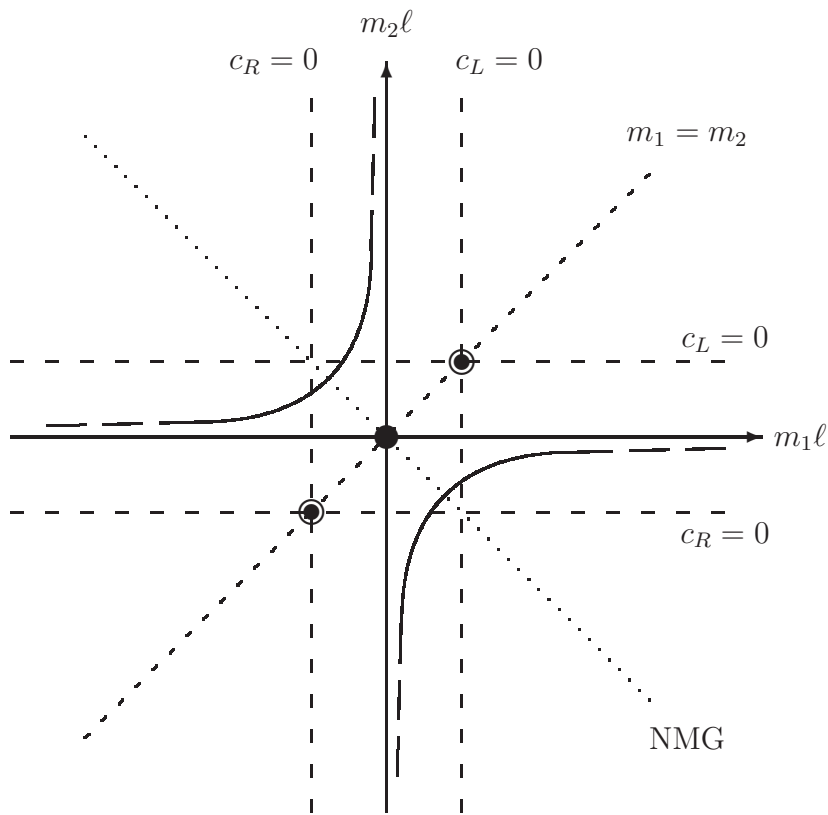

Figure \ref{fig:2} summarizes all critical loci in the parameter space of GMG. The two cases that are different from critical NMG are 1.~the dashed line where $m_1=m_2$ and 2.~the two double circles (``tricritical point'') where $m_1=m_2=\pm1/\ell$, leading to a rank-3 log CFT.
The full circle in the center is the partially massless point \cite{Bergshoeff:2009aq} (see \cite{Deser:1983mm,Deser:2001pe,Deser:2001us} for partial masslessness in AdS), where also log modes arise \cite{Grumiller:2010tj}.
The hyperbola corresponds to the scaling limit $m^2\to 0$.

We now review the form of the Jordan cell and the two-point functions of GMG at the tricritical point.
When acting with the Hamiltonian we obtain a rank-3 Jordan cell.
\begin{equation}
 H \left(\begin{matrix} \psi^{\rm log^2} \\ \psi^{\rm log} \\ \psi^L\end{matrix}\right) =
   2\left(\begin{matrix} 1 & 1 & 0 \\ 0 & 1 & 1 \\ 0 & 0 & 1 \end{matrix}\right)
   \left(\begin{matrix} \psi^{\rm log^2} \\ \psi^{\rm log} \\ \psi^L\end{matrix}\right) 
\end{equation}
This means, the left-moving part of the stress-energy tensor, $\psi^L$, has two partners, a log mode, $\psi^{\rm log}$, and a log-squared mode, $\psi^{\rm log^2}$. 
The two-point functions are completely determined by only one parameter $b_L$ which was calculated and labeled generalized
new anomaly in \cite{Grumiller:2010tj}. The correlators were calculated using similar tricks as in \cite{Grumiller:2009mw,Grumiller:2009sn}
and take the form
\begin{align}
 \langle \psi^L(z) \,\psi^L(0) \rangle &= \langle \psi^L(z) \,\psi^{\rm log}(0,0) \rangle =0 \label{eq:lcft41} \\
 \langle \psi^{L}(z) \,\psi^{\rm log^2}(0,0) \rangle &= \langle \psi^{\rm log}(z,\bar{z}) \,\psi^{\rm log}(0,0) \rangle = \frac{b_L}{2 z^4}  \label{eq:lcft42}  \\
 \langle \psi^{\rm log}(z,\bar{z}) \,\psi^{\rm log^2}(0,0) \rangle &= -\frac{b_L\,\ln|z|^2}{z^4} \label{eq:lcft43} \\
 \langle \psi^{\rm log^2}(z,\bar{z}) \,\psi^{\rm log^2}(0,0) \rangle &= \frac{b_L\,\ln^2|z|^2}{z^4}\,.  \label{eq:lcft44} 
\end{align}
The correlators \eqref{eq:lcft41}-\eqref{eq:lcft44} coincide precisely with corresponding rank-3 log CFT correlators \cite{Flohr:2001tj}. 
In principle every log-term in the correlators comes with a spurious mass parameter $m_L^2$, see (\ref{eq:lcftc3}) and the discussion afterwards. 
With no loss of generality we set $m_L^2=1$ in the correlators above.

Another characteristic feature of the tricritical theory is that the log$^2$-modes have an even weaker fall-off than log modes. This can be seen directly if we write the log$^2$-mode in the following form
\begin{equation}
 \psi^{\rm log^2}=4\big(it+\ln\cosh\rho\big)^2\,\psi^L \,. \label{eq:lcft68}
\end{equation}
In the large $\rho$ limit the asymptotic behavior of the log$^2$-mode differs from the left-moving primary by
a factor $\rho^2$. Weaker boundary conditions than Brown--Henneaux or log boundary conditions \eqref{eq:lcft36} have to be imposed to include this mode. Such boundary conditions were given in \cite{Liu:2009pha}.

Some further papers on the tricritical point in GMG are \cite{Bertin:2011jk,Kim:2012rz}.
Other theories that behave similarly to GMG at the linearized level were constructed, like theories with cubic and quartic curvature invariants \cite{Sinha:2010ai}, arbitrary powers of curvature invariants \cite{Paulos:2010ke} or Born--Infeld-like constructions \cite{Gullu:2010pc,Ghodsi:2011ua}.
AdS$_3$/log CFT$_2$ aspects of these theories were addressed in \cite{Grumiller:2010tj}.

\subsubsection{Parity-even tricritical gravity}

The study of GMG and its critical points exhausts the possible log CFTs that can be obtained from four-derivative theories.
Higher-derivative theories propagate more massive degrees of freedom in general, and can therefore lead to an even richer structure than GMG.
One such example is parity even tricritical (PET) gravity \cite{Bergshoeff:2012ev}.
\begin{multline}
 S_{\rm PET}=\frac{1}{16\pi G_N}\int{\rm d}^3x\sqrt{-g}\,\Big[\sigma R-2\Lambda_0
           -\Big(\frac{\al}{8\ell^2}+\frac{3\beta}{8}\Big)R^2+\beta R^{\mu\nu}R_{\mu\nu} \\
             +\al\Big(\nabla^\rho R^{\mu\nu}\nabla_\rho R_{\mu\nu}-\frac38\nabla_\mu R\nabla^\mu R\Big)\Big]
\end{multline}
The AdS radius is related to the parameter $\Lambda_0$ by $\Lambda_0=[\sigma-3\al/(4\ell^4)-\beta/(4\ell^2)](-1/\ell^2)$.
Much as NMG is the ``square'' of TMG --- in the sense that all chiral linearized solutions are doubled by acquiring a partner with opposite helicity --- PET gravity is the square of GMG. 

PET gravity has four massive helicity modes, apart from the massless Einstein modes. 
The linearized EOM are given by
\begin{equation}
 (\mathcal{D}^L\mathcal{D}^R\mathcal{D}^{m_1}\mathcal{D}^{m_2}\mathcal{D}^{m_3}\mathcal{D}^{m_4} \psi)_{\mu\nu}=0 
\end{equation}
where again $\mathcal{D}^{L/R}$ are exactly as in \eqref{eq:lcft30} and $\mathcal{D}^{m_i}$ are as in \eqref{eq:lcft31} with $\mu=m_i$.
The mass parameters are given by the four solutions of
\eq{
 (m_{1,2,3,4})^2=\frac{\beta}{2\al}+\frac{2}{\ell^2}\pm\frac{1}{2\al}\sqrt{\frac{10\al^2}{\ell^4}+\frac{6\al\beta}{\ell^2}+4\al\sigma+\beta^2}\,.
}{eq:lcft69}
All modes come in two pairs of opposite chirality. We have two free mass parameters, $m_1=-m_2$ and $m_3=-m_4$, associated with the two signs in \eqref{eq:lcft69}. 
We can achieve similar degenerations as for GMG, but now all dual log CFTs are parity even. 
There are three cases. 1.~Rank-2 log CFT with vanishing central charge by letting one of the masses go to zero,
       i.e.~$\mathcal{D}^{m_1}=\mathcal{D}^L$ [$(m_{1,2})^2=1/\ell^2\neq m_{3,4}^2$; this implies $\mathcal{D}^{m_2}=\mathcal{D}^R$].
2.~Rank-2 log CFT with non-zero central charge if the two massive modes degenerate
       $\mathcal{D}^{m_{1,2}}=\mathcal{D}^{m_{3,4}}$ [$m_{1,2}^2=m_{3,4}^2\neq1/\ell^2$].
3.~Rank-3 log CFT by tuning $\mathcal{D}^{m_1}=\mathcal{D}^{m_3}=\mathcal{D}^L$ [$(m_{1,2,3,4})^2=1/\ell^2$].
More on PET gravity can be found in \cite{Apolo:2012vv,Zojer:2012rj}.

\subsubsection{More critical models in three dimensions}

Including arbitrary numbers of derivatives in the gravity action allows to obtain Jordan cells \eqref{eq:lcft7} of arbitrary rank.
Gravity duals to higher-rank log CFTs were presented in \cite{Nutma:2012ss}. 
To obtain a rank-$n$ Jordan cell like \eqref{eq:lcft7} it is highly plausible that one has to further generalize the boundary conditions  from log boundary conditions
\eqref{eq:lcft36} to
\eq{
\ga_{ij} = \ga_{ij}^{(0)} e^{2\rho} + \sum_{m=1}^n \rho^m \ga_{ij}^{(1,m)} + \ga_{ij}^{(2)} + \dots 
}{eq:lcft45}
where the $\ga_{ij}^{(1,m)}$ additionally have to obey either some chirality conditions like \eqref{eq:chirality} or tracelessness conditions $\ga_{ij}^{(1,n)}\ga^{ij}_{(0)}=0$, depending on the specific theory. 
For $n=1$ and $n=2$ we know from the previous examples that \eqref{eq:lcft45} is correct, while for $n>2$ its correctness is very plausible.
This is so, because the factor between log- and left-modes in \eqref{eq:lcft35} scales asymptotically linearly in $\rho$; moreover, this factor has the right algebraic properties (see footnote \ref{fn:4}) to appear with the $m$-th power in the $m$-th log partner of the (anti-)holomorphic flux component of the stress-energy tensor; finally, the limiting construction analog to \eqref{eq:lcft14} or \eqref{eq:lcft35} is bound to produce such a factor.
Of course, it is useful (and straightforward) to confirm \eqref{eq:lcft45} for any given model. 
It could be of interest to consider gravity models that allow an $n\to\infty$ limit, see \cite{Rasmussen:2004jc}.

\section{Generalizations}\label{se:3}

In this section we provide some generalization of the AdS$_3$/log CFT$_2$ correspondence reviewed in the previous section.
In subsection \ref{se:3.0} we add supersymmetry and show how the superconformal log CFT structure emerges from critical 3-dimensional supergravity.
In subsection \ref{se:3.1} we consider higher-dimensional versions of the AdS/log CFT correspondence.
In subsection \ref{se:3.2} we generalize to different spins, including scalar fields and higher spin gravity, and mention also generalizations to other gravity backgrounds (e.g.~asymptotic Schr\"odinger).
In subsection \ref{se:3.3} we study \.In\"on\"u--Wigner contractions --- non- and ultra-relativistic limits of the 2-dimensional conformal algebra --- and provide some modest steps towards a contracted AdS/log GCA correspondence. We also clarify some minor lacuna in flat space chiral gravity concerning a rank-3 Jordan cell.

\subsection{Supersymmetry}\label{se:3.0}

Supersymmetric generalizations of log CFTs can be found via a similar limiting procedure as outlined in section \ref{se:1.2}. Here we focus on log CFTs that have the $\mathcal{N}=1$ super-Virasoro algebra as part of their symmetry algebra. This supersymmetric extension of the Virasoro algebra contains the usual Virasoro generators $L_n$, as well as a set of anti-commuting generators $G_n$. The (anti-)commutation relations are given by
\begin{align}
[L_n,L_m] &= (n-m)L_{n+m} + \frac{c_L}{12}\,(n^3-n)\,\delta_{n+m,0}  \\  
[L_n,G_m] &= (\tfrac{1}{2}n - m)G_{n+m} \\
\label{anticommutator}
\{G_n,G_m\} &= 2L_{n+m} + \frac{c_L}{3}\,(n^2-\tfrac{1}{4})\,\delta_{n+m,0}  \,.
\end{align}
A log CFT that has the above symmetries has two extra anti-commuting operators $S^L(z)$ and $S^{\mathrm{log}}(z)$, whose 2-point correlators can be obtained via the limiting procedure of section \ref{se:1.2}. The operator $S^L(z)$ has conformal weights $(h,\bar{h}) = (\tfrac32,0)$. One then again supposes that there is an additional operator $S_\varepsilon(z,\bar{z})$ with conformal weights $(h_\varepsilon,\bar{h}_\varepsilon) = (\tfrac32 + \varepsilon, \varepsilon)$, such that for finite $\varepsilon$ the following non-vanishing 2-point correlators are found.
\begin{align}
\langle S^L(z)\, S^L(0)\rangle & =  \frac{2 c_L}{3 z^3}  \\
\langle S_\varepsilon(z,\bar{z})\, S_\varepsilon(0,0)\rangle & =  \frac{c_\varepsilon}{2 z^{3+2\varepsilon} \bar{z}^{2 \varepsilon}} 
\end{align}
Defining the log operator $S^\mathrm{log}(z,\bar{z})$ via
\begin{equation}
S^\mathrm{log}(z,\bar{z}) = \lim_{\varepsilon \rightarrow 0} \frac{S_\varepsilon(z,\bar{z}) - S^L(z)}{\varepsilon} = \frac{\extd}{\extd \varepsilon} S_\varepsilon(z,\bar{z}) |_{\varepsilon = 0} \,,
\label{eq:lalapetz}
\end{equation}
yields the following correlators in the limit $\varepsilon \rightarrow 0$.
\begin{align}
\langle S^L(z)\, S^L(0)\rangle & =  0  \\
\langle S^{\mathrm{log}}(z,\bar{z}) S^L(0) \rangle & =  \frac{2 b_L}{3 z^3}  \\
\langle S^{\mathrm{log}}(z,\bar{z})S^{\mathrm{log}}(0,0)\rangle & =  - \frac{4 b_L \log(m_S^2 |z|^2)}{3 z^3} 
\end{align}
Analogously to the bosonic case \eqref{eq:lcftc3}, the spurious mass parameter $m_S$ can again be rescaled to any finite value by suitable shifts of the log mode \eqref{eq:lalapetz} by some finite amount of $S_L(z)$.

Supersymmetric versions of the AdS/log CFT correspondence can be found by looking at supersymmetric versions of critical gravity theories. Here we focus on the appearance of logarithmic modes in the supersymmetric version of cTMG, discussed in \cite{Becker:2009mk}.

Shortly after the inception of TMG, Deser and Kay constructed a supersymmetric version in \cite{Deser:1982sw}, which was later generalized by Deser to include a cosmological constant \cite{Deser:1982sv}. The latter is an $\mathcal{N}=(1,0)$ theory, as the supersymmetry parameter is a single Majorana spinor in the $(\mathbf{2},\mathbf{1})$ representation of the underlying $\mathrm{AdS}_3$ isometry group $SL(2) \times SL(2)$. The bulk action of super-TMG is given as a sum of an Einstein supergravity action and a super-Lorentz--Chern--Simons action
\begin{multline} \label{sugraTMG}
S =  \frac{1}{16 \pi G_N} \int \rmd^3 x \, e\,  \Big[ R + \frac{2}{\ell^2} - i \varepsilon^{\mu \nu \rho} \bar{\psi}_\mu \left(D_\nu(\omega) - \frac{1}{2\ell} \gamma_\nu \right) \psi_\rho  \\  -\frac{1}{2 \mu} \varepsilon^{\mu \nu \rho} \left( \partial_\mu \omega_\nu{}^{ab} \omega_{\rho b a} + \frac{2}{3} \omega_\mu{}^a{}_b \omega_\nu{}^b{}_c \omega_\rho{}^c{}_a  \right) + \frac{i}{2\mu} \bar{f}^\mu \gamma_\nu \gamma_\mu f^\nu \Big]
\end{multline}
where $f^\mu$ is the dual of the gravitino field strength.
\begin{equation}
f^\mu = \varepsilon^{\mu \rho \sigma} D_\rho(\omega) \psi_\sigma  \qquad \mathrm{where} \ \ \ D_\mu(\omega) \psi_\nu = \partial_\mu \psi_\nu + \frac14 \omega_\mu{}^{ab} \gamma_{ab} \psi_\nu 
\end{equation}
The action \eqref{sugraTMG} is given in  second-order form, i.e., the spin connection $\omega_\mu{}^{ab}$ is a dependent field, whose expression in terms of the vielbein and gravitino is given by the usual torsionfull spin connection of simple supergravity. The supersymmetry transformations that leave \eqref{sugraTMG} invariant, for all values of the coupling constants, are given by
\begin{align}
\delta e_\mu{}^a & =  i \bar{\epsilon} \gamma^a \psi_\mu   \\
\delta \psi_\mu & =  2 D_\mu(\omega) \epsilon - \frac{1}{\ell} \gamma_\mu \epsilon \,.
\end{align}
These transformations leave the first and second lines of \eqref{sugraTMG} separately invariant. Supersymmetry implies that, for non-singular values of the coupling constants, the massive graviton is accompanied by a local fermionic degree of freedom, the massive gravitino. 

Global AdS corresponds to a maximally supersymmetric background of $\mathcal{N}=(1,0)$ super-TMG, preserving an $\mathrm{OSp}(1|2;\mathbb{R})_L \times SL(2)_R$ (super-)group. In the dual boundary theory, this symmetry algebra is enhanced to a direct product of an ordinary Virasoro algebra in the anti-holomorphic sector and an $\mathcal{N}=1$ super-Virasoro algebra in the holomorphic sector. The values of the central charges are the same as the ones of bosonic TMG \eqref{eq:lcft28}. A supersymmetric version of chiral gravity, conjectured to be dual to an $\mathcal{N}=1$ chiral CFT, can thus be obtained at the same critical value of the coupling constants given in \eqref{eq:lcft29}. In cTMG, a linearized analysis shows the existence of a logarithmic mode \eqref{eq:lcft34}, which provides the basis for the AdS/log CFT correspondence for cTMG. In a similar way, a linearized analysis can be performed in critical super-TMG, to show that the logarithmic mode of cTMG is now accompanied by a logarithmic mode in the gravitino sector. The linearized equations of motion for the metric and gravitino decouple. For the metric, they are the same as for ordinary TMG and the analysis outlined in section \ref{se:2.1} can be repeated. For the gravitino, one finds that the equation of motion in gamma-traceless gauge, $\gamma^\mu \psi_\mu=0$, reduces to 
\begin{equation}
(\mathcal{D}^L \mathcal{D}^{(\mu)} \psi)_\nu = 0\,,
\end{equation}
where the mutually commuting operators $\mathcal{D}^L$, $\mathcal{D}^{(\mu)}$ are given by $\mathcal{D}^L = \mathcal{D}^{(\mu)}|_{\mu = 1/\ell}$ and 
\begin{equation}
\mathcal{D}^{(\mu)} = \slashed{\textfrak{D}} + \frac{1}{2\ell} - \mu \,.
\end{equation}
The covariant derivative $\textfrak{D}$ is with respect to the background spin connection and Levi-Civita connection. As for the graviton excitations, equations of the type $(\mathcal{D}^M \psi^M)_\mu = 0$ can be solved using a separation ansatz for the gravitino excitations
\begin{equation}
\psi^M_\mu = e^{-i h x^+ - i \bar{h} x^-} F_\mu(\rho) \left( \begin{array}{c} i \\ e^\rho \end{array} \right) 
\end{equation}
where $h$, $\bar{h}$ correspond to the weights with respect to the $L_0$ and $\bar{L}_0$ AdS Killing vectors. As for the graviton modes, for each value of $\mu$ there exists a unique $\psi^M$ that is regular, normalizable and primary and whose explicit expression can be found in \cite{Becker:2009mk}. For positive $\mu$, the weights of this gravitino primary are given by 
\begin{equation}
(h,\bar{h}) = (1+\frac{\mu \ell}{2},\,-\frac12 + \frac{\mu \ell}{2}) \,.
\end{equation}
This means the primary $\psi^L$ annihilated by $\mathcal{D}^L$ has weights $(\tfrac32,\,0)$. For $\mu \ell =1$, the operators $\mathcal{D}^L$ and  $\mathcal{D}^{(\mu)}$ and the weights of primaries annihilated by them degenerate with each other. As for the graviton modes, this signals the existence of a gravitino log mode that obeys
\begin{equation}
(\mathcal{D}^L \mathcal{D}^L \psi^\mathrm{log} )_\mu = 0  \qquad (\mathcal{D}^L \psi^\mathrm{log})_\mu \propto \psi^L_\mu \,.
\end{equation}
This superpartner of the graviton log mode can again be constructed via a limiting procedure (with $\varepsilon = \frac{\mu \ell-1}{2}$):
\begin{equation}
\psi^\mathrm{log}_\mu = \lim_{\varepsilon \rightarrow 0} \frac{\psi_\mu^M - \psi_\mu^L}{\varepsilon} = -2 (i t + \ln \cosh \rho) \psi^L_\mu \,.
\label{eq:lcft78}
\end{equation} 
The prefactor on the right hand side is precisely the same one as in the bosonic case \eqref{eq:lcft35}.
The Jordan cell of the bosonic sector is now accompanied by a similar Jordan cell for the Hamiltonian $H$ in the gravitino sector,
\begin{equation}
H \begin{pmatrix} \psi_\mu^{\mathrm{log}} \\ \psi_\mu^L \end{pmatrix}  = 2 \begin{pmatrix} \frac34 & 1 \\ 0 & \frac34 \end{pmatrix}  \begin{pmatrix} \psi_\mu^{\mathrm{log}} \\ \psi_\mu^L \end{pmatrix}  
\end{equation}
whereas the angular momentum operator $J$ is still diagonal.
In conclusion, supersymmetrization of the AdS/log CFT correspondences seems straightforward.

\subsection{Generalization to higher dimensions, AdS$_{d+1}$/log CFT$_{d}$}\label{se:3.1}

The past years have also seen efforts to generalize the intriguing correspondence of critical gravity theories with log CFTs to higher dimensions, closely following the ideas employed previously in the 3-dimensional literature.
For this reason we will merely state a few selected results that show the close similarity to the 3-dimensional
case that we discussed already at length in section \ref{se:2}. 

As 3-dimensional CFTs are drastically different from their lower-dimensional cousins, so are their logarithmic counterparts. The main difference is that the symmetry algebra is finite, which is common to all CFTs in more than two dimensions. We focus our review on the AdS$_4$/log CFT$_3$ correspondence and list higher-dimensional work at the end of this subsection.

L\"u and Pope introduced 4-dimensional ``Critical Gravity'' \cite{Lu:2011zk}. 
The bulk action is given by
\begin{equation}
 S_{CG} = \frac{1}{16\pi G_N}\,\int{\rm d}^4x\sqrt{-g}\, \big[R-2\Lambda+\beta\big(R^2-3R_{\mu\nu}R^{\mu\nu}\big)\big] \,.
\end{equation}
The critical tuning is obtained as $\beta=-1/(2\Lambda)=\ell^2/6$. 
Linearization around an AdS background, together with the transversal gauge choice $\nabla_\mu(\psi^{\mu\nu}-g^{\mu\nu}\psi^\lambda{}_\lambda)=0$, where $\psi_{\mu\nu}$ is the perturbation, yields linearized EOM of the following form:
\begin{equation}\label{critgraveom}
\Big(\Box-\frac{2\Lambda}{3}\Big)\Big(\Box-\frac{4\Lambda}{3}-\frac{1}{3\beta}\Big)\psi_{\mu\nu}=0
\end{equation}
The two differential operators in the linearized EOM \eqref{critgraveom} degenerate when $\beta=-1/(2\Lambda)$. 
Subsequent work showed the existence of (linearized and non-linear) logarithmic solutions
\cite{Alishahiha:2011yb,Gullu:2011sj,Bergshoeff:2011ri,Chen:2011in,Gurses:2012db,AyonBeato:2012da} at the critical
point. In \cite{Porrati:2011ku} an inner product on the space of solutions was introduced and it was shown that
critical gravity with log solutions is not a unitary theory.

The truncation of the logarithmic solutions was discussed as well, and it was argued that Critical Gravity with Brown--Henneaux-like boundary conditions is trivial \cite{Lu:2011ks}. 
This is related to Maldacena's proposal to obtain Einstein gravity from 4-dimensional conformal gravity by imposing suitable boundary conditions \cite{Maldacena:2011mk}. 
Namely, truncated Critical Gravity is trivial, because its cosmological Einstein and Weyl-squared parts effectively describe the same (IR) physics, thus mutually canceling each others non-trivial excitations \cite{Lu:2011ks}.

We exhibit now the Jordan cell structure that arises in 4-dimensional Critical Gravity.
In \cite{Bergshoeff:2011ri} the linearized solutions were categorized for spin-2 using group-theoretic techniques. 
Denoting the Einstein solution by $\psi_{\mu\nu}$ the logarithmic solution is given by 
\begin{equation}
 \psi_{\mu\nu}^{\rm log}=-2\big(it+\ln\cosh\rho\big) \psi_{\mu\nu} \,.
\label{eq:angelinajolie}
\end{equation}
The factor appearing on the right hand side of \eqref{eq:angelinajolie} coincides precisely with the corresponding factor in 3 dimensions, see \eqref{eq:lcft35}.
The modes $\psi_{\mu\nu}^{\rm log}$ and $\psi_{\mu\nu}$ are the two independent solutions to the EOM
(\ref{critgraveom}) at the critical point $\beta=-1/(2\Lambda)$. 
One can show that $\psi_{\mu\nu}$ is an eigenfunction of the 4-dimensional analogs of the 3-dimensional Hamiltonian and spin operators $H$ and $J$ \cite{Bergshoeff:2011ri}, but $\psi_{\mu\nu}^{\rm log}$ is a generalized eigenfunction. 
Under the action of $H$ we find
\begin{equation}\label{critgravcell}
 H \left(\begin{matrix} \psi^{\rm log}_{\mu\nu} \\ \psi_{\mu\nu} \end{matrix}\right) =
    2\left(\begin{matrix} \tfrac32 & 1 \\ 0 & \tfrac32  \end{matrix}\right)
    \left(\begin{matrix} \psi^{\rm log}_{\mu\nu} \\ \psi_{\mu\nu} \end{matrix}\right) 
\end{equation}
while both functions are again eigenfunctions of $J$, see \eqref{eq:lcft15}. 
This shows the appearance of a Jordan cell in Critical Gravity.

Two-point functions in Critical Gravity were calculated using holographic techniques. 
In \cite{Johansson:2012fs} the correlators of the stress-energy tensor $T_{ij}$ (sourced by the non-normalizable Einstein modes $\psi_{ij}$) and its logarithmic partner $t_{ij}$ (sourced by the non-normalizable log modes proportional to $\psi^{\rm log}_{ij}$) were found to match the expectations for a rank-2 logarithmic pair. 
They are given by
\begin{align}
 \langle T_{ij}(x) T_{kl}(0) \rangle &= 0  \label{eq:lcft79}\\
 \langle T_{ij}(x) t_{kl}(0) \rangle &=
       \frac{b}{(2\pi)^3}\,\hat{\Delta}_{ij,kl}\,\frac{1}{|x|^2}  \label{eq:lcft80} \\
 \langle t_{ij}(x) t_{kl}(0) \rangle &=
       \frac{b}{(2\pi)^3}\,\hat{\Delta}_{ij,kl}\,\frac{\ln(m^2|x|^2)}{|x|^2}  \label{eq:lcft81}
\end{align}
with the ``new anomaly''
\eq{
b = \frac{3\ell^2}{4 G_N}
}{eq:lcft73}
and $\hat{\Delta}_{ij,kl}=\tfrac12\,(\hat\Theta_{ik}\hat\Theta_{jl}+\hat\Theta_{il}\hat\Theta_{jk}-\hat\Theta_{ij}\hat\Theta_{kl})$ where $\hat\Theta_{ij}=\partial_i\partial_j-\de_{ij}\,\square$.
As expected the two-point function between two stress-energy tensors vanishes and with one logarithmic partner
insertion it takes the form of the two-point function of two stress-energy tensors in an ordinary CFT (\ref{eq:lcft80}). The correlator between two logarithmic operators is similar, but contains a log-term.
The correlators \eqref{eq:lcft79}-\eqref{eq:lcft81} generalize the log CFT correlators \eqref{eq:lcftc1}-\eqref{eq:lcftc3} to higher dimensions.

Most of the above mentioned literature already covers critical gravity in arbitrary dimensions. 
More work can be found in \cite{Deser:2011xc,AyonBeato:2011qw,Sisman:2011gz,Kleinschmidt:2012rs}.

\subsection{Generalizations to different spins and/or backgrounds}\label{se:3.2}

We summarize here some work on log gravity that differs from the main theme of this review, critical gravity in an AdS background and its holographic correspondence to a log CFT with spin-2 logarithmic partners, in one way or another. 

Let us start with changing the spin of the logarithmic partners to the simpler case of spin-0.
Gravity duals to log CFTs have been constructed using scalar field toy models already in the late 1990ies, see \cite{Ghezelbash:1998rj,Kogan:1999bn,Myung:1999nd,Lewis:1999qv}. 
Using a singleton dipole action
\eq{
S \sim \int\extd^3x\sqrt{-g}\,\Big(-g^{\mu\nu}\partial_\mu\Phi\partial_\nu\Phi^{\rm log} - m^2 \Phi\Phi^{\rm log} + \tfrac12\,\Phi^2\Big)
}{eq:lcft74}
leads to fourth order field equations on an AdS background
\eq{
(\nabla^2-m^2)\,\Phi=0\qquad (\nabla^2-m^2)\,\Phi^{\rm log}=\Phi\qquad  (\nabla^2-m^2)^2\,\Phi^{\rm log}=0
}{eq:lcft75}
that exhibit $\Phi$ and $\Phi^{\rm log}$ as logarithmic pair and eventually yield scalar 2-point correlators with a structure similar to \eqref{eq:lcftc1}-\eqref{eq:lcftc3}.
More recent papers are \cite{Bergshoeff:2012sc,Moon:2012vc,Moon:2012wr}.

The spin-1 case apparently has not received comparable attention yet.
Topologically massive gauge theories \cite{Deser:1982vy,Deser:1982wh,Deser:1982a} lead to similar behavior of the linearized excitations as TMG, see \cite{Carlip:2008jk}.
Therefore, critical tuning should lead to a Jordan-cell structure in the spin-1 sector and to a logarithmic  partner of the photon (or gluon).
It might be rewarding to analyze the spin-1 case analog to the spin-2 case reviewed in this work or to the spin-0 case mentioned in the previous paragraph, in particular to demonstrate the appearance of a Jordan cell for critical tuning, to calculate correlators and to consider holographic applications of logarithmic gauge field pairs.
Some papers on topologically massive electrodynamics and generalizations thereof where the issue of Jordan cells could be of interest are \cite{Minces:1999tp,Yee:2011yn,Fosco:2011ra,Andrade:2011sx} (see also references therein).

Given the recent interest in (3- or higher-dimensional) higher-spin gravity, see e.g.~\cite{Henneaux:2010xg,Campoleoni:2010zq,Gaberdiel:2012uj,Vasiliev:2012vf,Ammon:2012wc} and references therein, it may be worthwhile to recover Jordan cells in a higher-spin context.
This is not necessarily straightforward beyond the linearized level. 
In dimensions higher than 3 it seems far too difficult to envisage some TMG- or NMG-like higher derivative higher-spin theory.
Even in 3 dimensions the construction of such actions is not straightforward; indeed, in most of the literature exclusively the Chern--Simons formulation of higher-spin gravity \cite{Henneaux:2010xg,Campoleoni:2010zq} is considered, where the geometric part of the theory has no local physical degrees of freedom, just like in Einstein gravity \cite{Achucarro:1986vz,Witten:1988hc} or conformal Chern--Simons gravity \cite{Horne:1988jf}.
In order to obtain logarithmic partners for higher spin fields one has to go beyond the Chern--Simons formulation and consider some higher-spin analog of, say, TMG.
A higher-spin analog of TMG was introduced in \cite{Chen:2011vp,Bagchi:2011vr}.
Some further aspects like the 1-loop partition function \cite{Bagchi:2011td} and quasi-normal modes \cite{Kim:2012pt} where studied at the critical point.
On general grounds, if rank-2 Jordan cells appear in AdS$_3$ higher-spin gravity they should be of the form
\eq{
H\begin{pmatrix}
  \psi^{\rm log} \\ \psi
 \end{pmatrix}
= 2\,\begin{pmatrix}
   \frac{s}{2} & 1 \\ 0 & \frac{s}{2}
  \end{pmatrix}
 \begin{pmatrix}
  \psi^{\rm log} \\ \psi
 \end{pmatrix}
}{eq:lcft77}
where $H=L_0+\bar L_0$ and $s$ is the spin of the mode $\psi$ with weights $(s,\,0)$.
The angular momentum operator is again expected to be diagonal, with eigenvalues $s,s$.
On the higher spin gravity side the log modes $\psi^{\rm log}_{\mu_1 \dots \mu_s}$ again should be equal to their partner $\psi_{\mu_1 \dots \mu_s}$, multiplied by the same factor as in \eqref{eq:lcft35}, \eqref{eq:lcft78} or \eqref{eq:angelinajolie}.

Apart from changing the spin, one can also change the background geometry and thereby the asymptotic symmetries.
This can lead to gauge/gravity correspondences that also involve critical points, degeneration of modes, Jordan cells etc., though not necessarily in a log CFT context.
Critical gravity theories on background geometries different from AdS were studied in
\cite{AyonBeato:2011qw,Bergshoeff:2011xy,Setare:2012zv,Hyun:2012fd}.

\subsection{Inonu--Wigner contractions and log Galilean Conformal Algebras}\label{se:3.3}

\.In\"on\"u--Wigner contractions allow to construct new (symmetry) algebras from given ones by suitable rescalings and limits of the generators.
The best known \.In\"on\"u--Wigner contraction is the one from the (inhomogeneous) Lorentz group to the Galilei group as the speed of light is sent to infinity \cite{Inonu:1953}.
More recently, Bagchi and Gopakumar considered the contraction of the relativistic conformal algebra to the Galilean Conformal Algebra (GCA) in an attempt to construct the non-relativistic limit of AdS/CFT \cite{Bagchi:2009my}, and showed that this could be given an infinite lift for all spacetime dimensions. In two dimensions, the infinite dimensional GCA can be constructed out of a contraction of two copies of the Virasoro algebra \cite{Bagchi:2009pe}. 
As with the Schr\"odinger algebra \cite{Son:2008ye,Balasubramanian:2008dm}, the GCA has been considered in the context of non-relativistic holography. See \cite{Duval:2009vt,Martelli:2009uc,Alishahiha:2009np,Bagchi:2009ca,Sakaguchi:2009de,Bagchi:2009pe} for some applications and elaborations. 
Interestingly, the 2-dimensional GCA is also relevant for discussions in the context of flat-space holography where it emerges from an ultra-relativistic limit as the asymptotic symmetry algebra at null infinity \cite{Barnich:2012aw,Bagchi:2012yk,Barnich:2012xq,Bagchi:2012xr,Barnich:2012rz}.\footnote{%
Through the BMS/GCA correspondence \cite{Bagchi:2010zz,Bagchi:2012cy} the more recent work is related to pioneering papers by Barnich and collaborators on the BMS$_3$ algebra, see for instance \cite{Barnich:2006av,Barnich:2010eb,Barnich:2011mi}, which is isomorphic to GCA.
}
Given that log CFTs can arise as limiting cases of ordinary CFTs (see section \ref{se:1.2}) and that GCAs  arise as limiting cases of conformal algebras it is natural to combine both limits and construct log GCAs. 
In this review we are specifically interested in log GCAs and their holographic duals.

In two dimensions a GCA is obtained either from a non-relativistic  or an ultra-relativistic limit of the conformal algebra.
For the time being we focus on the former.
Given two Virasoro algebras with generators ${\cal L}_n$, $\bar{\cal L}_n$ and central charges $c_L$, $c_R$ one can define rescaled linear combinations
\eq{
L_n = {\cal L}_n + \bar {\cal L}_n\qquad M_n = \de \big({\cal L}_n - \bar {\cal L}_n\big)
}{eq:lcft46}
where $\de$ is a small parameter that eventually we send to zero.
Defining the central charges (subscripts $V,T$ refer to ``Virasoro'' and ``Translations'')
\eq{
c_V = \lim_{\de\to 0} \big(c_L + c_R\big) \qquad c_T = \lim_{\de\to 0} \big(\de (c_L-c_R)\big)
}{eq:lcft47}
one obtains then the centrally extended GCA \cite{Bagchi:2009my}.
\begin{align}
[L_n,\,L_m]&=(n-m)\,L_{n+m}+\frac{c_V}{12}\,(n^3-n)\,\de_{n+m,\,0} \label{eq:lcft48} \\
[L_n,\,M_m]&=(n-m)\,M_{n+m}+\frac{c_T}{12}\,(n^3-n)\,\de_{n+m,\,0} \label{eq:lcft49} \\
[M_n,\,M_m]&=0 \label{eq:lcft50} 
\end{align}
The GCA vacuum is defined by the conditions $L_n|0\rangle=M_n|0\rangle=0$ for $n>-2$ and connects continuously to the $sl(2)$ invariant CFT vacuum defined by ${\cal L}_n|0\rangle=\bar{\cal L}_n|0\rangle=0$ for $n>-2$. 
The light-cone coordinates $z,\bar z$ in the non-relativistic limit are rescaled to
\eq{
z \to  t + \de x \qquad \bar z \to t - \de x\,.
}{eq:lcft51}
The flux components of the stress-energy tensor $\psi^{L/R}$ combine to the operators $\psi^V$ and $\psi^T$.
\eq{
\psi^V (t,x) = \lim_{\de\to 0} \big[\psi^L(z) + \psi^R(\bar z)\big]\qquad \psi^T(t,x) = \lim_{\de\to 0} \big[\de\big(\psi^L(z) - \psi^R(\bar z)\big)\big]
}{eq:lcft52}
Their 2-point functions, after taking the limit $\de\to 0$, take the simple form 
\begin{align}
 \langle \psi^T(t,x) \psi^T(0,0) \rangle &= 0 \label{eq:lcft53} \\ 
 \langle \psi^T(t,x) \psi^V(0,0) \rangle &= \frac{c_T}{2t^4} \label{eq:lcft54} \\
 \langle \psi^V (t,x) \psi^V (0,0) \rangle &= \frac{c_V}{2t^4} - \frac{2c_T x}{t^5} \,. \label{eq:lcft55}
\end{align}
Interestingly, even without a log mode there is a Jordan cell in a GCA,
\eq{
M_0 \begin{pmatrix}\psi^V \\ \psi^T\end{pmatrix}=2\begin{pmatrix}0 & 1 \\ 0 & 0\end{pmatrix} \begin{pmatrix}\psi^V \\ \psi^T\end{pmatrix}
}{eq:lcft70}
while $L_0$ is diagonal with eigenvalues $2,2$.
However, since we do not have a relativistic CFT after contraction the presence of a Jordan cell does not imply the existence of logarithms in correlation functions.
Still, there is structural similarity to log CFTs in the correlators: $\psi^T$ has vanishing correlator with itself but non-vanishing one with $\psi^V$, while the $\psi^V$ correlator has a more complicated form.

The ultra-relativistic limit works analogously, though there are subtle differences to the non-relativistic limit.
Defining
\eq{
L_n = {\cal L}_n - \bar {\cal L}_{-n}\qquad M_n = \de \big({\cal L}_n + \bar {\cal L}_{-n}\big)
}{eq:lcft61}
and
\eq{
c_V = \lim_{\de\to 0} \big(c_L - c_R\big) \qquad c_T = \lim_{\de\to 0} \big(\de (c_L+c_R)\big)
}{eq:lcft62}
yields again the GCA \eqref{eq:lcft48}-\eqref{eq:lcft50}.
Note, however, that the GCA vacuum is not connected to the standard $sl(2)$ invariant CFT vacuum, but instead  emerges from a CFT vacuum with the conditions ${\cal L}_n|0\rangle=\bar{\cal L}_{-n}|0\rangle=0$ for $n>-2$.
The minus sign in the index of the bar-sector essentially reverses the sign of the right-moving central charge, which explains the sign changes in \eqref{eq:lcft62} as compared to \eqref{eq:lcft47}, and similar sign changes in the formulas below.
The light-cone coordinates $z,\bar z$ in the ultra-relativistic limit are rescaled to
\eq{
z \to  \de t + x \qquad \bar z \to \de t - x\,.
}{eq:lcft63}
The flux components of the stress-energy tensor $\psi^{L/R}$ combine to
\eq{
\psi^V (t,x) = \lim_{\de\to 0} \big[\psi^L(z) -\psi^{R\dagger}(-\bar z)\big]\qquad \psi^T(t,x) = \lim_{\de\to 0} \big[\de\big(\psi^L(z) + \psi^{R\dagger}(-\bar z)\big)\big]\,.
}{eq:lcft64}
In the ultra-relativistic limit the 2-point correlators of the GCA are identical to the correlators \eqref{eq:lcft53}-\eqref{eq:lcft55} with the replacement $t\leftrightarrow x$.
For the calculations below we note that the original Virasoro zero mode generators act on the GCA states $\psi^{V/T}$ as follows.
\eq{
{\cal L}_0 \psi^V = 2 \psi^L(z)\quad \bar{\cal L}_0 \psi^V = 2 \psi^{R\dagger}(-\bar z)\quad {\cal L}_0 \psi^T = 2 \de \psi^L(z)\quad \bar{\cal L}_0 \psi^T = -2 \de\psi^{R\dagger}(-\bar z)
}{eq:lcft76}

We consider now log GCAs.
To define a log GCA we demand two properties, 1.~the presence of a Jordan cell in either $L_0$ or $M_0$ (or both) and 2.~the presence of a logarithm of either $t^2$ or $x^2$ in at least one of the 2-point correlators.
In the ultra-relativistic limit it is straightforward to repeat the construction of section \ref{se:1.2} and add a massive operator $\psi_\eps$ that degenerates with $\psi^L$ in the limit of vanishing $\eps$, so that again a log operator \eqref{eq:lcft13} can be defined.
Since we require $c_L\to 0$ in the limit while keeping finite $c_V$, we automatically obtain $c_T\to 0$.
In this way we obtain the non-vanishing 2-point correlators of an ultra-relativistic log GCA.
\begin{align}
 \langle \psi^V (t,x) \psi^V (0,0) \rangle &= \frac{c_V}{2x^4} 
\label{eq:lcft58} \\
 \langle \psi^{\rm log} (t,x) \psi^V (0,0) \rangle &= \frac{b_L}{2x^4}  \label{eq:lcft59} \\
 \langle \psi^{\rm log} (t,x) \psi^{\rm log} (0,0) \rangle &= -\frac{b_L\ln{(m_L^2x^2)}}{2x^4}  \label{eq:lcft60}
\end{align}
The quantity $m_L$ can again be changed by redefining the log modes.
Shifts $\psi^{\rm log}\to(1-2\ga)\tilde\psi^{\rm log}+\ga\psi^V$ permit to eliminate the mixed correlator \eqref{eq:lcft59} for a suitable choice of the parameter $\ga$. 
This redefinition leaves the Jordan cell structure invariant, see \eqref{eq:lcft72} below. 
At the same value for $\ga$ the mass parameter $m_L$ is extremal.
For cTMG we obtain the following results for the coefficients in the correlators \eqref{eq:lcft58} and \eqref{eq:lcft60}
\eq{
c_V=-\frac{3\ell}{G_N}\qquad b_L=-\frac{3\ell}{G_N}\,. 
}{eq:lcft56}
While $L_0$ is diagonal with eigenvalues $2,2,2$, the generator $M_0$ acquires a Jordan cell.
\eq{
M_0 \begin{pmatrix} \psi^{\rm log} \\ \psi^V \\ \psi^T \end{pmatrix} = \begin{pmatrix}0 & 0 & 1 \\ 0 & 0 & 2 \\ 0 & 0 & 0 \end{pmatrix} \begin{pmatrix} \psi^{\rm log} \\ \psi^V \\ \psi^T \end{pmatrix}
}{eq:lcft72}

Even though there were important pre-cursors, the log GCA algebra \eqref{eq:lcft58}-\eqref{eq:lcft60} had not appeared yet in the literature to the best of our knowledge.
In \cite{Bagchi:2010vw} correlation functions in TMG were studied which coincide with the ones above [or rather with \eqref{eq:lcft53}-\eqref{eq:lcft55}], but without considering the log operator.
In \cite{Hosseiny:2010sj} log correlators in log GCAs were considered, which were then specified to cTMG in \cite{Hosseiny:2011ct}; however, their setup differs from the one above.
Note that $\psi^V$ has non-vanishing correlator with itself, in contrast to the situation encountered in \eqref{eq:lcftc1}; on the other hand, in cTMG $\psi^T$ has vanishing correlators with all operators.
In the non-relativistic limit considered by Hosseiny and Naseh \cite{Hosseiny:2011ct} we were not able to make sense of a log GCA.
This is so, because the log operator \eqref{eq:lcft13} behaves singularly when acting with $L_0$ from \eqref{eq:lcft46}.
\eq{
L_0 \psi^{\rm log} = \frac{(2+\eps)\psi^\eps - 2\psi^L}{\eps} + \frac{\eps\psi^\eps}{\eps} = 2\psi^{\rm log} + 2\psi^L = 2\psi^{\rm log} + \psi^V + \frac1\delta \psi^T
}{eq:lcft71}
We took first the limit $\eps\to 0$ and see that we can no longer take the limit $\de\to 0$ on the right hand side of \eqref{eq:lcft71}.
This differs drastically from the ultra-relativistic limit, where $\psi^{\rm log}$ is an eigenstate of $L_0$.
Note that the troubling issues with \eqref{eq:lcft71} could be avoided by defining massive modes at finite $\eps$ with weights $(2+\eps,\,-\eps)$. However, on the gravity side such modes would not have an interpretation as massive graviton excitations, since their spin, $h-\bar h=2+2\eps$, would not equal to two at finite $\eps$. 
It is possible that reversing the order of limits or taking simultaneously $\eps,\de\to 0$ while keeping fixed their ratio leads to other log GCA-like theories \cite{Hosseiny:2011ct}, in particular to cases where $c_T\neq 0$ is allowed, but a full understanding of all possibilities is currently lacking.

Frequently GCAs are studied as flat-space limit, $\ell\to\infty$, of the asymptotic AdS symmetry algebra.
Considering such a limit in cTMG requires to simultaneously send $\mu\to 0$, to maintain the condition \eqref{eq:lcft29}, and $G_N\to\infty$, to obtain finite central charges \eqref{eq:lcft56}.
This is precisely the scaling limit that leads to conformal Chern--Simons gravity \cite{Afshar:2011yh,Afshar:2011qw}.
Therefore, conformal Chern--Simons gravity with flat-space boundary conditions could lead to a log GCA.
This is potentially worrying, since the recent proposal \cite{Bagchi:2012yk} suggests that conformal Chern--Simons gravity with flat-space boundary conditions should be dual to a unitary (chiral) CFT, whereas log GCAs necessarily are non-unitary.

We show now how this issue is resolved.
In transverse-traceless gauge the linearized EOM of conformal Chern--Simons gravity on a flat-space background simplify to
\eq{
\big(({\cal D}^{0})^3\psi\big)_{\mu\nu}=0
}{eq:lcft66} 
where ${\cal D}^{0}$ is the operator ${\cal D}^{(\mu)}$ from \eqref{eq:lcft31} for $\mu=0$.
This implies actually a double degeneration: the ``massive'' graviton degenerates not only with the left-moving boundary graviton, but also left- and right-moving boundary gravitons degenerate with each other.
(This is possible only in the $\ell\to\infty$ limit.)
Taking the flat-space limit of \eqref{eq:lcft35} and \eqref{eq:lcft68} we should therefore expect a log mode $\psi^{\rm log}$ and a log-squared mode $\psi^{\rm log^2}$, related to their partner $\psi$ by
\eq{
\psi^{\rm log}=-2it \psi \qquad \psi^{\rm log^2}=-4t^2\psi \,.
}{eq:lcft67}
Together they should form a rank-3 Jordan cell with respect to the operator $M_0$ (but not with respect to $L_0$).
This is indeed precisely what happens \cite{Bagchi:2012yk}.
However, there is a catch: the modes $\psi$ (and its m{\'e}nage-{\`a}-trois-partners $\psi^{\rm log}$ and $\psi^{\rm log^2}$) are either singular or incompatible with the boundary conditions.
Nevertheless, in flat-space chiral gravity there are regular modes $\psi^{\rm reg}$ that solve the linearized EOM \cite{Bagchi:2012yk}.
What happens is that they cannot be represented in transverse-traceless gauge, as this gauge choice turns out to be singular in the flat-space limit.
We show now that the (singular) gauge transformation that maps the singular modes $\psi$ onto regular modes $\psi^{\rm reg}$ does {\em not} map any linear combinations of singular modes $\psi^{\rm log}$ and $\psi^{\rm log^2}$ to regular versions thereof.
Actually, all the modes must be pure gauge from a bulk perspective, so we can write $\psi=\nabla_{(\mu}\xi_{\nu)}$, $\psi^{\rm reg}=\nabla_{(\mu}\xi^{\rm reg}_{\nu)}$, and similarly for $\psi^{\rm log}$ and $\psi^{\rm log^2}$. 
The gauge transformation that maps $\psi$ to $\psi^{\rm reg}$ is therefore generated by the vector field $\xi^{\rm reg}-\xi$.
Acting with this vector field (by means of Lie-derivative) on linear combinations of $\psi^{\rm log}$ and $\psi^{\rm log^2}$ then always leads to modes that are either singular or disobey the boundary conditions (we checked this explicitly using the results for the modes $\psi$, $\psi^{\rm reg}$, $\psi^{\rm log}$ and $\psi^{\rm log^2}$ provided in \cite{Bagchi:2012yk}).
In conclusion, the flat-space boundary conditions of \cite{Bagchi:2012yk} automatically eliminate the log and log-squared modes, so that no log GCA arises and unitarity is maintained in flat-space chiral gravity.

\section{Loose ends}\label{se:4}

In this section we collect some loose ends, but make no claim to be complete.
In subsection \ref{se:4.2} we address condensed matter applications of log CFTs.
In subsection \ref{se:4.1} we list some missing checks of the AdS/log CFT correspondence, many of which are straightforward.
In subsection \ref{se:4.3} we speculate about some open issues that require more thought and/or novel ideas.

\subsection{Condensed matter applications}\label{se:4.2}

So far the AdS/log CFT correspondence has been used exclusively in one direction, namely to exploit knowledge about log CFTs to better understand the corresponding critical gravity theories.
The purpose of this subsection is twofold: 1.~to express the hope that AdS/log CFT could be useful also in the other direction --- namely, to exploit knowledge about critical gravity theories and various probes therein to better understand the corresponding log CFTs --- and 2.~to list applications of log CFTs that have shown up in the literature over the past two decades.

Log CFTs have appeared in a number of different contexts in condensed matter physics and related subjects, like 2-dimensional turbulence \cite{Flohr:1996ik,RahimiTabar:1996ki,Skoulakis:1998} (see also \cite{Polyakov:1992er,Bernard:2007,Fedorenko:2012zc}), abelian sandpile models \cite{Mahieu:2001,Ruelle:2002jy,Jeng:2004ea,Piroux:2004vd,Piroux:2004si,MoghimiAraghi:2004wg,Jeng:2006tg,Saberi:2008zr,Caracciolo:2008pu,Poghosyan:2010ts,Caracciolo:2011ab}, critical polymers \cite{Saleur:1991hk,Flohr:1995ea,Kausch:1995py,Ivashkevich:1998na,Cardy:1999zp,Pearce:2006we,Pearce:2009us,Pearce:2012bu}, D-brane recoil \cite{Kogan:1996zv,Ellis:1996fi,Mavromatos:1998nz,Ellis:1999jf,Ellis:1999sf,Ellis:2000sx} or decay \cite{Lambert:2003zr}, fractional quantum-Hall effect \cite{Flohr:1995zj,Gurarie:1997dw,Cappelli:1998ma,Ino:1998xe}, gravitational dressing \cite{Bilal:1995rc}, percolation \cite{Mathieu:2007pe,Rasmussen:2007pc,Kytola:2008uz,Rasmussen:2008ii,Ridout:2008cv,Vasseur:2012tz},  symplectic fermions \cite{Kausch:2000fu}, and systems with disorder \cite{Caux:1995nm,Kogan:1996wk,Maassarani:1996jn,Caux:1996kq,Caux:1998sm,Guruswamy:1999hi,Ludwig:2000em,Bhaseen:2000,RahimiTabar:2001et} or quenched disorder \cite{Cardy:1999zp,Gurarie:1999yx,RezaRahimiTabar:2000qr,Cardy:2001,Gurarie:2004ce,Dubail:2010zz}.
(For further references see \cite{Flohr:2001zs,Gaberdiel:2001tr}.)

We focus here on systems with quenched disorder, as this allows to apply the type of $c=0$ log CFT that arises naturally on the gravity side.  
The discussion below is a condensed version of section 5 in \cite{Grumiller:2010rm}, to which we refer for details.
Different tools have been developed to study systems with quenched disorder, including ``supersymmetric methods'', replica field theory and log CFTs. 
Recently, Fujita, Hikida, Ryu and Takayanagi combined the replica method with the AdS/CFT correspondence to describe disordered systems \cite{Fujita:2008rs} (see \cite{Kiritsis:2008at,Myers:2008me,Adams:2011rj,Adams:2012yi} for related work), essentially by taking $n$ copies of the CFT, exploiting AdS/CFT to calculate correlators and taking formally the limit $n\to 0$ in the end. Like other replica tricks their approach relies on the existence of the limit $n\to 0$.
One of the results obtained by the supersymmetric method or replica trick is that correlators develop a logarithmic behavior, exactly as in a log CFT \cite{Cardy:1999zp}. In fact, in the $n\to 0$ limit prescribed by the replica trick, the conformal dimensions of certain operators degenerate. This produces a Jordan block structure for the Hamiltonian, the defining feature of a log CFT.

It is then suggestive to exploit the AdS/log CFT correspondence reviewed in the present work to holographically describe systems with quenched disorder.
Actually, any $c=0$ log CFT where the stress-energy tensor acquires a logarithmic partner is a candidate for a theory with a holographic critical gravity dual. 
Currently no such applications exist in the literature.
It could be rewarding to study such applications, essentially by lifting the numerous results on AdS/CFT to results applicable in AdS/log CFT.
For reviews on applications of the gauge/gravity correspondence to condensed matter physics see \cite{McGreevy:2009xe,Sachdev:2011wg,Iqbal:2011ae}.

A holographic description of log CFTs with $c\neq 0$ is possible as well, see section \ref{se:2.3}.
Moreover, generalizations to higher dimensions (summarized in section \ref{se:3.1}) could be useful to holographically study higher-dimensional log CFTs, where the literature is rather scarce so far both on the gravity and the CFT side.

\subsection{Missing checks}\label{se:4.1}

We list now further possible checks of the AdS/log CFT correspondence. 
\begin{itemize}
  \item {\bf Boundary conditions.} The most straightforward check is to verify the boundary conditions \eqref{eq:lcft45} (or higher-dimensional analogs thereof) for critical gravity theories with Jordan cells the rank of which is bigger than three. Consistency of the variational principle as well as integrability, finiteness and conservation of the canonical charges can then be verified for any given theory.
  \item {\bf Supersymmetry.} 
One can also wonder whether  versions of the AdS/log CFT correspondence with extended supersymmetry can be found. This would require the construction of higher-derivative gravity theories, like TMG and NMG, that exhibit extended supersymmetry. Unlike the $\mathcal{N}=1$ case, where super-GMG has been constructed at the fully non-linear level \cite{Andringa:2009yc,Bergshoeff:2010mf}, results concerning $\mathcal{N}>1$ versions of TMG, NMG and GMG are only known at the linearized level, around Minkowski space \cite{Bergshoeff:2010ui}. It was shown that linearized NMG can have up to $\mathcal{N}=8$ supersymmetry, whereas parity-violating models like TMG can only have up to $\mathcal{N}=7$ supersymmetry (which may be further reduced to $\mathcal{N}=6$ non-linearly). For AdS/log CFT applications fully non-linear versions of these theories are needed, which have not been constructed yet.
  \item {\bf Holographic renormalization.} As a pre-requisite to some of the checks below holographic renormalization of the (super-)gravity action is necessary. For instance, the construction of suitable boundary counterterms for supersymmetric GMG is still lacking. In 3-dimensional theories it should be possible to construct them along the lines of \cite{Grumiller:2009dx}, i.e., by demanding supersymmetry without imposing any boundary conditions on the fields \cite{Belyaev:2005rt,Belyaev:2007bg,Belyaev:2008xk}. Alternatively, one can straightforwardly generalize the results by Skenderis, Taylor and van Rees \cite{Skenderis:2009nt}.
 \item {\bf Correlators.} Critical TMG so far is the only critical gravity theory where 3-point correlators have been calculated on the gravity side \cite{Grumiller:2009mw}. It could be of interest to perform similar checks for other critical gravity theories in various dimensions. Actually, even in cTMG 3-point correlators that involve two log insertions and a left insertion or three log insertions have not been calculated completely yet, but only in the large weight limit (and without extracting the residues from the poles in the weights).
More accessible missing checks are the verification of the conformal Ward identities for critical gravity theories with higher rank Jordan cells, like tricritical GMG or PET gravity (see section \ref{se:2.3}), and the superconformal Ward identities for supersymmetric cTMG (see section \ref{se:3.0}) or supersymmetric extensions of higher-dimensional Critical Gravity \cite{Lu:2011mw}.
It would be nice to fill these gaps.
 Unlike 2- and 3-point correlators, 4-point correlators (or higher) are not restricted uniquely by the conformal Ward identities, but instead involve some free functions. The simplest examples of this type are 4-point correlators on the gravity side. This, however, will involve lengthy calculations, starting with the fourth variation of some higher derivative action, like cTMG, critical NMG, or some of the generalizations discussed in sections \ref{se:2.3} and \ref{se:3}. 
 \item {\bf Classical partition functions.} Some of the checks involve the calculation of partition functions (see sections \ref{se:1.3} and  \ref{se:2.2}). In the semi-classical approximation the leading contribution comes from the classical part. For theories without Chern--Simons terms, like Einstein gravity [see \eqref{eq:lcft23}] or NMG the classical partition function is derived easily from the on-shell action, see for instance \cite{Papadimitriou:2005ii}. However, for TMG this method fails. Currently, the only way to derive the classical partition function for TMG is to calculate its canonical charges \cite{Deser:2003vh}, derive entropy \cite{Kraus:2005vz,Solodukhin:2005ah,Tachikawa:2006sz}, impose the validity of the first law and integrate the latter. A disadvantage of this method is that the first law is not derived, but rather an input. It would be nice to find a method to derive the free energy of BTZ black holes in TMG that does not require the validity of the first law.
 \item {\bf Quantum partition functions.} Checks analogous to the ones in critical TMG or NMG \cite{Gaberdiel:2010xv} (see section \ref{se:2.2}) are still lacking for all higher-dimensional generalizations of critical gravity theories. Moreover, a better understanding of the 1-loop partition function from the log CFT side is desirable, in particular the combinatorics of multi-log excitations. Finally, in general it is not true that the partition function is 1-loop exact, so that calculations of 2- or higher-loop partition functions could be necessary.
\end{itemize}

\subsection{Open issues}\label{se:4.3}

We conclude with some intriguing open issues.
\begin{itemize}
 \item {\bf Log modes and black holes.} It is not known whether BTZ black holes can be created dynamically in cTMG, e.g.~from the collapse of log modes. This is quite different from the situation in pure Einstein gravity with a scalar field, which can collapse to a BTZ black hole \cite{Pretorius:2000yu,Husain:2000vm}. It was indicated by Deser and Franklin \cite{Deser:2010df} that BTZ black holes belong to a separate superselection sector, but no formal proof of this statement exists yet. The stability analysis in \cite{Birmingham:2010gj} may be taken as another hint that BTZ black holes and log modes do not talk to each other. Finally, higher-derivative theories like cTMG also allow black hole solutions that are not locally AdS. It could be interesting to verify if there is any qualitative change in the behavior of log modes in such backgrounds as compared to BTZ.
 \item {\bf Log entanglement entropy.} Entanglement entropy is holographically related to minimal surfaces in  AdS$_3$ \cite{Ryu:2006bv,Ryu:2006ef} and, like the Bekenstein--Hawking entropy, it is proportional to the central charge in a CFT$_2$ \cite{Holzhey:1994we,Vidal:2002rm,Calabrese:2004eu}. (Another feature shared between Bekenstein--Hawking and entanglement entropies is the area law \cite{Bombelli:1986rw,Srednicki:1993im}, see also \cite{Eisert:2008ur}.) Entanglement entropy therefore has become another useful holographic tool. It would be of interest to consider entanglement entropy in the context of the AdS/log CFT correspondence. 
 \item {\bf Log transport coefficients.} Holography is efficient in calculating transport coefficients at strong coupling. An example is the shear viscosity calculated by virtue of the Kubo formula from 2-point correlators of the stress-energy tensor \cite{Policastro:2001yc,Kovtun:2004de}. Within the AdS$_{d+1}$/log CFT$_d$ correspondence ($d\geq 3$) it could be possible to define analogs of shear viscosity that involve ``Kubo formulas'' with 2-point correlators containing the log partner of the stress-energy tensor.  
 \item {\bf Log GCAs.} We have provided a working definition of log GCAs and a limiting procedure to construct them from a 1-parameter family of ordinary CFTs in section \ref{se:3.3}, but more general limits might be possible, perhaps along the lines of \cite{Hosseiny:2011ct}.
A better understanding of the structure of these theories is desirable.
 \item {\bf Log higher spin gravity.} In section \ref{se:3.2} we addressed briefly generalizations to log modes with spins different from 2. Except for the spin-0 case very little is known about such theories, both on the gravity side and the CFT side. Perhaps a more manageable warm-up exercise is to consider the spin-1 case, which is conspicuously absent in the current literature.
 \item {\bf Log excitations on non-AdS backgrounds.} A number of holographic correspondences beyond AdS/CFT were suggested in the past five years. Typically, they involve highly (but not maximally) symmetric gravity backgrounds, like asymptotically warped \cite{Anninos:2008fx,Anninos:2009zi,Compere:2009zj}, Schr\"odinger \cite{Son:2008ye,Balasubramanian:2008dm}, Lifshitz \cite{Kachru:2008yh} or Lobachevsky \cite{Bertin:2012qw}. All these backgrounds arise in TMG (see \cite{Ertl:2010dh} for all stationary axi-symmetric solutions and \cite{Chow:2009km,Chow:2009vt} for a summary of exact solutions in TMG), so one does not have to consider very exotic bulk actions to obtain such backgrounds. It is conceivable that in some of these gauge/gravity correspondences there exist critical tunings where two or more of the operators in the dual field theory degenerate with each other, and a log CFT-like structure with Jordan cells emerges. On the gravity side a possible first step is to comprehensively discuss the spectrum of linearized excitations around these backgrounds and search for possible degeneracies of the modes at some critical tunings.
 \item {\bf Log Cardy formula?} The Cardy formula has been useful in microscopic derivations of black hole entropy \cite{Strominger:1997eq}. Is there some analog of the Cardy formula for log CFTs, and if so, does it involve the ``new anomaly''?
 \item {\bf Logs from multiple M2 branes?} It is possible that cTMG is dual to a log CFT, as conjectured in \cite{Grumiller:2008qz}. Alternatively, cTMG could be merely some effective theory that is dual to a log CFT in a certain limit --- much like the supergravity approximation in the standard AdS/CFT correspondence is valid only at large $N$ and for large 't~Hooft coupling. In that case it would be very interesting to find a string-completion of the AdS/log CFT correspondence. This is a challenge not just from a technical viewpoint, but also conceptually, because string theory is a unitary theory, whereas log CFTs are non-unitary. Intriguingly, Chu, Nilsson and collaborators found evidence \cite{Chu:2009gi,Chu:2010fk} for the emergence of cTMG in the context of multiple M2 branes \cite{Bagger:2006sk,Gustavsson:2007vu,Bagger:2007jr,Bagger:2007vi,Aharony:2008ug}, namely  in the higgsed phase of conformal supergravity coupled to ABJM Chern--Simons matter, see also \cite{Nilsson:2012ky}.
It is currently unknown what is the fate of the log modes in these models and whether a Jordan cell exists.
If the log modes are present then (topologically gauged) multiple M2 branes could provide a string-completion of cTMG and the holographic AdS$_3$/log CFT$_2$ correspondence.
 \item {\bf Gravity dual of log Ising model?} We have focussed so far mostly on log CFTs whose central charge vanishes and where the stress-energy tensor acquires a logarithmic partner. However, from the field theory point of view there are much simpler log CFTs available, like the seminal $c=-2$ example \cite{Gurarie:1993xq,Flohr:1995ea,Kausch:1995py,Gaberdiel:1996np,Rohsiepe:1996qj,Flohr:2001zs,Gaberdiel:2001tr,Fjelstad:2002ei,Pearce:2006sz,Read:2007qq} or log minimal models \cite{Pearce:2006sz}, including the ``log Ising model''. Recently, evidence was provided that Einstein gravity with central charge $c=\tfrac12$ is dual to the Ising model \cite{Castro:2011zq}. If this turns out to be true, then it seems natural to speculate what is the gravity dual of the $c=-2$ model or the gravity dual of the log Ising model. 
\end{itemize}

\section*{Acknowledgments}

We thank Uwe G\"unther and David Ridout for the kind suggestion to prepare this review article. 
DG is indebted to Roman Jackiw and Niklas Johansson for their pioneering contributions to and collaboration on the story reviewed in the present work.
We are grateful to Hamid Afshar, Joao Aparicio, Arjun Bagchi, Eric Bergshoeff, Mario Bertin, Steve Carlip, Alejandra Castro, Branislav Cvetkovic, Stanley Deser, Stephane Detournay, Sjoerd de Haan, Sabine Ertl, Matthias Gaberdiel, Michael Gary, Parsa Ghorbani, Olaf Hohm, Roman Jackiw, Niklas Johansson, Finn Larsen, Esperanza Lopez, Robert McNees, Wout Merbis, Ali Naseh, Ioannis Papadimitriou, Radoslav Rashkov, Ivo Sachs, Stefan Stricker, Peter van Nieuwenhuizen and Dima Vassilevich for collaborations and numerous colleagues for discussions during the past five years, in particular the participants of the 2009 ESI workshop ``Gravity in three dimensions'' in Vienna, the 2011 IHP program ``Advanced Conformal Field Theory and Applications'' in Paris, the 2012 ESI workshop ``Higher Spin Gravity'' and the 2012 KITP program ``Bits, Branes and Black holes'' in Santa Barbara. 
DG thanks John McGreevy for helpful remarks about log CFTs at a critical point in the early discussions on cTMG in May 2008 at MIT.
We are grateful to Arjun Bagchi and Stephane Detournay for numerous useful comments, in particular concerning section \ref{se:3.3}.

This work was supported by the START project Y 435-N16 and projects P 21927-N16, I 952-N16 and I 1030-N27 of the Austrian Science Fund (FWF).
TZ was supported by a grant of the Dutch Academy of Sciences (KNAW).



\section*{References}


\providecommand{\href}[2]{#2}\begingroup\raggedright\endgroup

\end{document}